# Light-induced temperature jump causes power-dependent ultrafast kinetics of electrons generated in multiphoton ionization of liquid water.[1]

Robert A. Crowell, *[a] Rui Lian, [a] Ilya A. Shkrob, [a]
Jun Qian, [a,2)] Dmitri A. Oulianov, [a] and Stanislas Pommeret [a,b]

[a] *Chemistry Division, Argonne National Laboratory, 9700 S. Cass Ave., Argonne, IL 60439.*
[b] *CEA/Saclay, DSM/DRECAM/SCM/URA 331 CNRS 91191 Gif-Sur-Yvette Cedex, France.*

**Abstract**

Picosecond geminate recombination kinetics for electrons generated by multiphoton ionization of liquid water become power dependent when the radiance of the excitation light is greater than 0.3-0.5 TW/cm$^2$ (the terawatt regime). To elucidate the mechanism of this power dependence, tri- 400 nm photon ionization of water has been studied using pump-probe laser spectroscopy on the pico- and femtosecond time scales. We suggest that the observed kinetic transformations are caused by a rapid temperature jump in the sample. Such a jump is inherent to multiphoton ionization in the terawatt regime, when the absorption of the pump light along the optical path becomes very nonuniform. The heating of water is substantial (tens of $^o$C) because the 3-photon quantum yield of the ionization is relatively low, ca. 0.42, and a large fraction of the excitation energy is released into the solvent bulk as heat. Evidence of the temperature jump is the observation of a red shift in the absorption spectrum of (thermalized) electron and by characteristic "flattening" of the thermalization dynamics in the near IR. The temperature jump in the terawatt regime might be ubiquitous in multiphoton ionization in molecular liquids. The implications of these observations for femtosecond pulse radiolysis of water are discussed.

___

[*] To whom correspondence should be addressed: *Tel* 630-252-8089, *FAX* 630-2524993, *e-mail:* rob_crowell@anl.gov.

[2] presently at the Experimental facility Division, Advanced Photon Source, Argonne National Laboratory, 9700 S. Cass Ave., Argonne IL 60439; *Tel* 630-252-5874, *FAX* 630-2529303, *e-mail:* jqian@aps.anl.gov.

# 1. Introduction.

Multiphoton ionization of liquid water exhibits two distinct energetic regimes that are determined by the total excitation energy: When this total energy is below a certain threshold value (ca. 9.5 eV), the spatial distribution of the ejected photoelectrons ($e_{aq}^-$) around their parent holes is narrow (ca. 1 nm in width) and the geminate recombination of these electrons is relatively fast. Above this threshold energy, the distribution increases to 2-3 nm (ca. 12.4 eV) and the geminate recombination kinetics becomes relatively slow. [1,2,3] It is believed that the photoelectron ejected in the high-energy ionization process is promoted directly into the conduction band of the liquid, whereas in the low energy photoionization, some other process occurs, because the energy is insufficient to access this band. [1,3] This (autoionization) process should involve a proton transfer in concert with the loss of the electron; the latter is sometimes regarded as an electron transfer to a "pre-existing trap" in the bulk water. [4,5,6] The width of the initial electron distribution is then defined by the availability of suitable electron traps. [5,6] Several models for this low-energy photoprocess have been suggested (see refs. 1 and 3 for more detail); the important point is that in all of these models, such a photoprocess is viewed entirely differently from the high-energy ionization.

Interestingly, the transformation of a narrow electron distribution (that manifests itself through rapid geminate kinetics and low survival probability for the electron) into a broad electron distribution (that manifests itself through slow geminate kinetics and greater survival probability) can be brought about simply by changing the radiance of the pump light. [1,7,8,9] Such changes have been observed by several ultrafast spectroscopy groups, including this laboratory. In particular, Crowell and Bartels [1,2] demonstrated that rapid geminate kinetics of hydrated electrons observed in three 400 nm photon ionization of water becomes slower and the fraction of escaped electrons increases when the radiance of the 400 nm light is in the terawatt range. The same trend was observed in the similar terawatt regime by Schwartz et al. [7] and Eisenthal et al. [8] What causes such a dramatic transformation?

Crowell and Bartels suggested that this changeover is caused by the absorption of one or more 400 nm photons by a light-absorbing species generated within the duration of the short excitation pulse (which, in their experiments, was 4-5 ps fwhm). [1,2] In the following, this mechanism (which may be regarded as the condensed-phase analog of the above threshold ionization in the gas phase) is referred to as the "3+1" mechanism. The suggestion was made that the species in question is a pre-thermalized or thermalized (solvated) electron generated in the course of photoionization. Subsequent studies by Barbara and coworkers [10] seemed to support such a scenario: it was shown that one- 400 nm photon excitation of fully thermalized, hydrated electron promotes it into the conduction band of the solvent where it thermalizes and localizes after migrating > 1 nm away from the hole. Consequently, the geminate recombination is suppressed and the decay of the electron is slowed down. This photoprocess is very efficient: the quantum yield of the suppression is near unity. Thus, one of the possible mechanisms for the "3+1" photoprocess is the absorption within the duration of the ionizing pulse of a 400 nm photon by the electrons generated in the photoionization process; the electron gains sufficient energy to penetrate deeply into the conduction band of the solvent, reaching the same extended states that are accessed in the course of high-energy water ionization. This "3+1" mechanism naturally explains why the resulting electron distribution is broad.

While such a mechanism seems plausible for picosecond photoionization of water, it is less clear how such a mechanism can account for the changeover in the kinetic behavior observed for shorter, femtosecond pulses. The thermalization of the electron takes a relatively long time (up to 2 ps), whereas the pre-thermalized electron is a poor light absorber in the blue. In fact, the spectral manifestation of the thermalization/localization process is a rapid blue shift of the absorption band of the electron initially centered in the near IR to the visible; [11-15] the initial absorbance of pre-thermalized electron at 400 nm is negligible. [11,13,14] Even fully thermalized, hydrated electron has relatively low absorptivity at 400 nm (the molar absorptivity is ca. 2600 $M^{-1}$ $cm^{-1}$). [16] Since the photoexcitation of the electron competes with that of the solvent itself, and absorption of 400 nm light by water is very efficient in the terawatt regime, it seems unlikely that a large fraction of photoelectrons can be excited within the duration of a short excitation pulse. The estimates given in the Appendix (see the Supporting

Information), using the photophysical parameters determined in section 3.1 of the present work, confirms these suspicions: even in the most favorable scenario, too few electrons would be photoexcited by 400 nm photons, and the observed transition from one kinetic regime to another could not occur. Thus, the only way to explain the observed kinetics changeover using the "3+1" mechanism is to postulate that some other species, perhaps a short-lived excited state of water, absorbs the extra 400 nm photon. However, this explanation is presently not supported by our data since no light-absorbing state with the postulated properties has been observed on the sub-picosecond time scale (section 3.2). Either this 400 nm light-absorbing species has extremely short lifetime (and then it cannot absorb a sufficiently large number of 400 nm photons to make the "3+1" mechanism efficient) or such a species (e.g., the excited state) does not exist. In the latter case, it becomes unclear what species is responsible for the "3+1" photoexcitation.

In this article, the terawatt regime of 400 nm photoionization of water is reinvestigated on the femto- and pico- second time scales. We argue that the changeover in the kinetic profiles observed in the terawatt regime is caused by a rapid temperature jump, which is inherent to multiphoton ionization in the terawatt regime. The heating of water is substantial (tens of $^{\circ}C$) because the 3-photon quantum yield of the ionization is relatively low, ca. 0.42, and a large fraction of the excitation energy is released into the solvent bulk as heat. In this terawatt regime, the absorption of light is extremely nonuniform (for excitation light radiance > 1 TW/cm$^2$, most of the 400 nm photons are absorbed in a 10 μm thick layer near the surface; see section 3.1), and the local density of the ionization and heat-release events is very high. Above a certain critical concentration (achieved at ca. 1 TW/cm$^2$) the released heat rapidly (< 5 ps) reaches the geminate pair, and the geminate recombination thereby occurs in hot water. Evidence of the temperature jump is the observation of a red shift in the spectrum of (thermalized) electron and by characteristic "flattening" of the thermalization dynamics in the near IR. The latter effect has also been observed in bi- 266 nm photonic ionization of water in the terawatt regime. [9,17] For this 266 nm ionization, the 2-photon quantum yield is also relatively low, ca. 0.52. [17a] It is possible that the temperature jump is a general feature of multiphoton ionization of molecular liquids in the terawatt regime.

To save space, the Appendix and some figures (Figs. 1S to 8S) are given in the Supporting Information.

**2. Experimental.**

*Laser spectroscopy.* Transient absorbance (TA) measurements reported below were obtained using a 1 kHz Ti:sapphire setup the details of which are given in ref. 18. A diode-pumped Nd:YVO laser was used to pump a Kerr lens mode-locked Ti:sapphire laser operating at 80 MHz (Spectra Physics Tsunami). The 45 fs fwhm pulses from the oscillator were stretched to 80 ps in a single grating stretcher. Single pulses were then selected at 1 kHz with a Pockels cell. The 2 nJ pulses were amplified to 4 mJ in a home-built two-stage multipass Ti:sapphire amplifier. The amplified pulses were passed through a grating compressor that yielded Gaussian probe pulses of 60 fs fwhm and 3 mJ centered at 800 nm. The pulse-to-pulse stability was typically 3%.

The amplified beam was split into two parts. One beam was used to generate the probe pulses while the other was used to generate the 400 nm (second harmonic) pump pulses by frequency doubling in a thin BBO crystal. The pump power was determined using a calibrated thermopile meter (Ophir Optronics model 2A-SH). The probe pulse was derived from the same Ti:sapphire laser beam and delayed in time as much as 600 ps using a 3-pass 25 mm delay line. For kinetic measurements, 150-200 points acquired on a quasi-logarithmic grid were typically used. If not stated otherwise, the probe pulse was 800 nm. To obtain probe light of wavelength $\lambda$ other than 800 nm, a white light supercontinuum was generated by focussing the 800 nm light on a 1 mm thick sapphire disk; the probe light was selected using 10 nm fwhm interference filters.

The pump and probe beams were perpendicularly polarized, focused to round spots and overlapped in the sample at $5^o$. The spot sizes were determined by scanning the beam with a 10 μm pinhole or by imaging on a CCD array. The maximum fluence of the 400 nm photons was 0.4 J/cm$^2$. At the high end of this power range, dielectric breakdown of water was observed.

Fast Si photodiodes were used to measure the transmitted and the reference probe and pump light for every laser shot. Testing these photodiodes against the thermopile detector as a function of 400 nm light radiance showed perfect linearity of the photoresponse over the entire power range. The photodiode signals were amplified and sampled using home-built electronics and a 16-bit A/D converter. A mechanical chopper locked at 50% repetition rate of the laser was used to block the pump pulses on alternative shots. The standard deviation for a pump-probe measurement was typically 10 µOD, and the "noise" was dominated by the variation of the pump intensity and flow instabilities in the liquid jet.

In some kinetic measurements, the radius of the pump beam was 2-3 times smaller than the radius of the probe beam (usually, the pump beam envelops the probe beam at the sample). This "reverse" geometry was used to reduce the sensitivity of the TA signal to imperfections in the 400 nm beam and thermal lensing and to minimize the long-term walk-off of the two beams relative to each other as the delay stage is moved. Naturally, the TA signal obtained with the "reverse" beam geometry is much smaller than that obtained using the "normal" beam geometry (pump beam enveloping the probe beam at the sample), and its power dependence is different.

The transmission of 400 nm light $T_p$, the pump radiance $J_p$, and photoinduced absorbance $\Delta OD_\lambda$ at the probe wavelength $\lambda$ were determined from these measurements. To obtain the power dependencies, 20-50 series of $(1-5) \times 10^3$ pulses were averaged; the vertical and horizontal bars given in the plots correspond to 95% confidence limits using Pearson analysis. The pump power was continually monitored, and shots in which the absolute deviation of the power exceeded 1-3% of the mean value were rejected; the same protocol was used to obtain the TA kinetics. The typical rejection rate was 10-50%, depending on the pump power.

The pulse widths of the pump and probe pulses were determined by harmonic generation in a thin BBO crystal. Alternatively, the rise time of the TA signal from free photocarriers generated by 400 nm light in a 1.4 µm thick film of amorphous Si:H alloy (8 at. % H) was determined. [19]

*Materials and the flow system.* The experiments were carried out with a home-made high-speed (10 m/s) jet with a stainless steel 90 µm nozzle. In some experiments, a better-quality jet with a 70 µm sapphire nozzle (V. Kyburz AG) was used. To obtain the kinetics in the low-radiance limit, the kinetic measurements were carried out using (i) a home-made 560 μm thick high-speed jet or (ii) a 5 mm optical path cell with suprasil windows (in the latter case, the pump beam was focussed in the center of the cell to avoid damage to the windows). Pure He- or $N_2$-saturated water with dc conductivity < 2 nS/cm was circulated by a gear pump in an all 316 stainless steel and Teflon flow system. The experiments above the room temperature were carried out as explained in ref. 18.

**3. Results and Discussion.**

*3.1. Power dependence of pump transmission and product yield.*

We first consider an idealized photosystem in which the absorbance of the pump light by the photoproduct(s) is negligible (see ref. 17a for the derivation of equations). A more complex situation in which the photoproduct (e.g., the electron) also absorbs the pump light is considered in the Appendix (see the Supplement). Let $I(z,r;t)$ be the time-dependent amplitude

$$I(z,r;t) = J_p \, \exp\left(-r^2/\rho_p^2 - t^2/\tau_p^2\right) \tag{1}$$

of the axially-symmetric, plane-wave Gaussian laser pulse propagating in the direction $z$ through the medium with $n$-photon absorption coefficient $\beta_n$ (here $z=0$ corresponds to the incidence at the sample). In eq. (1), $\rho_p$ and $\tau_p$ are the $1/e$ beam radius and pulse width of the Gaussian pulse, respectively, and $J_p$ is the (peak) pump radiance (so that the total pump power $I_p$ is given by $I_p = \pi^{3/2} J_p \rho_p^2 \tau_p$). The absorption of the pump light by the sample is given by equation

$$\partial I(z,r;t)/\partial z = - \beta_n \, I(z,r;t)^n, \tag{2}$$

and the formation rate of the product (i.e., the electron) generated by simultaneous absorption of $n$ photons with quantum yield of $\phi_n$ is given by

$$\partial c(z,r;t)/\partial t = - \phi_n/n \, \partial I(z,r;t)/\partial z, \tag{3}$$

where $c(r,z;t)$ is the concentration of the photoproduct. We will assume that this photoproduct does not absorb the pump light during the laser excitation of water and does not decay in the time interval before detection. From eq. (3) we obtain that the optical density of the photoproduct by the end of the photoexcitation pulse is given by

$$\beta_{pr} \int_0^L dz \ c(z,r;t=+\infty) \ = \ \phi_n \beta_{pr}/n \int_{-\infty}^{+\infty} dt \ [I(0,r;t) - I(L,r;t)], \tag{4}$$

where $L$ is the sample thickness and $\beta_{pr}$ is the absorption coefficient of the product at the probe wavelength. For $n > 1$, eq. (2) has the solution

$$I(z,r;t) \ = \ I(0,r,t)/\left(1 + \beta_n z \ (n-1) \ I(0,r,t)^{n-1}\right)^{1/(n-1)}. \tag{5}$$

Combining eqs. (1) to (5), the transmission of the pump light $T_p$ through the sample is given by

$$T_p = \left(2/\sqrt{\pi} J_p \rho_p^2 \tau_p\right) \int_0^\infty dr \ r \int_{-\infty}^{+\infty} dt \ I(L,r;t), \tag{6}$$

and the photoinduced change $\Delta T_{pr}$ in the transmission $T_{pr}$ of the probe light is given by

$$-\Delta T_{pr} = \frac{2}{\rho_{pr}^2} \int_0^\infty dr \ r \ \exp\left(-\frac{r^2}{\rho_{pr}^2} - \frac{\phi_n \beta_e}{n} \int_{-\infty}^{+\infty} dt \ [I(0,r;t) - I(L,r;t)]\right) \tag{7}$$

where $I(L,r;t)$ is given by eq. (5) for $z = L$ and $\rho_{pr}$ is the $1/e$ radius of the Gaussian probe beam. The photoinduced change $\Delta OD$ in the decadic optical density of the sample (i.e., the TA signal) is related to the quantity given by eq. (7) by $\Delta OD = -\log(1 + \Delta T_{pr}/T_{pr})$. Using eqs. (6) and (7), the absorption coefficient $\beta_n$ and the quantum yield $\phi_n$ can be determined for the known parameters $\rho_p$, $\tau_p$, and $\rho_{pr}$ from the dependencies of $\Delta OD$ and $T_p$ vs. $J_p$ (Fig. 1). For $\rho_{pr} \ll \rho_p$, only the beam of light with $r \approx 0$ can be considered, and the averaging over the polar coordinate in eq. (7) is unnecessary. In such a case, the function $C(z) = c(z,r=0;t=+\infty)$ gives the concentration profile of the reaction product along the path of the probe light (Fig. 2(a)). [17a,b] The higher the pump radiance, the steeper is this profile. Using these concentration profiles, the decay kinetics of the electron due to cross recombination in the bulk can be simulated (Fig. 2(b) and section 3.4)

In Fig. 1, a typical dependence of the TA signal from the electron for $\lambda$=800 nm probe light (trace (i)) and 400 nm light transmittance of the pump laser pulse (trace (ii)) vs. the incident pump radiance is shown by symbols. The TA signal of the (hydrated) electron along the path of 800 nm light traversing the $L$=70 μm thick water jet was determined at the delay time of 5 ps, when the thermalization dynamics is over and the geminate recombination of the electron is still negligible (section 3.4). For this measurement, the pump pulse was 189 fs fwhm ($\tau_p$=120 fs) and the pump and probe beams were 210 μm fwhm ($\rho_p$=126 μm) and 40 μm fwhm ($\rho_{pr}$=24 μm), respectively. The refractive index $n_w$ of water of 1.343 for 405 nm light obtained in ref. 20 was used to calculate the reflectivity of the jet surface, $R$=0.0214 (multiple reflections were neglected), and the pump radiance used in eq. (1) was corrected by a factor of $(1-R)$. The decadic molar absorptivity of 18500 M$^{-1}$ cm$^{-1}$ for hydrated electron was used. [16] The solid lines in Fig. 1 are least-squares fits to eqs. (6) and (7) assuming a 3-photon photoprocess ($n$ = 3) which gave the optimum parameters $\beta_3$=267±4 cm$^3$/TW$^2$ and $\phi_3$=0.415±0.006. The absorption coefficient $\beta_3$ obtained in this study is considerably lower than 900 cm$^3$/TW$^2$ given by Naskrecki et al. [20] who used a thicker, 1 cm optical path sample and low-radiance 400 nm light (< 0.1 TW/cm$^2$) for water photoexcitation. Such disagreements are fairly typical for measurements of photophysical parameters for multiphoton excitation. [17a]

As seen from the linear section of the plot of $\Delta OD_{800}$ vs. $J_p$ given in Fig. 1, the power law, $\Delta OD_{800} \propto J_p^3$, holds to $J_p$ < 0.3 W/cm$^3$. Above this radiance, the TA curve bends down; in the same range, the transmission of 400 nm light rapidly decreases with increase in the laser power. The bending over of the photoinduced absorption is due to the nonuniformity of the 400 nm light absorption by the jet; the typical (simulated) concentration profile $C(z)$ of the photoelectrons along the beam axis for $J_p$=1.6 TW/cm$^2$ is shown in Fig. 2(a), trace (i). Compare this profile with the concentration profiles simulated for lower radiance, $J_p$=1 TW/cm$^2$ (Fig. 2(a), trace (ii)) and $J_p$=0.5 TW/cm$^2$ (Fig. 2(a), trace (iii)). It is seen that in the terawatt regime, most of the 400 nm light is absorbed in a thin 10 μm layer near the jet surface. This shortening of the effective optical path of the probe light is the reason for slowing down (seen in Fig. 1) of the increase in the $\Delta OD_{800}$ with increasing $J_p$. When a straight line is drawn on this double logarithmic plot through the data points for which $J_p > 0.5$ TW/cm$^2$ (trace (iii) in Fig. 1), the slope of this line is close to unity, and this behavior initially led us to believe that "3+1" photoexcitation is vigorous in this terawatt regime. A detailed model for this "3+1" photoprocess is given in the Appendix and Figs. 1S, 2S, and 3S in the Supplement. As seen from Fig. 2S therein, the postulated absorption of 400 nm photons by photoproducts (e.g., electrons) is not needed to explain the power dependencies shown in Fig. 1, insofar as the inhomogeneous light absorption alone explains the observed trend. Furthermore, as shown in the Appendix, when the occurrence of the "3+1" photoexcitation is postulated, the fit quality in Fig. 1 does not improve.

While the general behavior shown in Fig. 1 was reproducible, the power dependencies of the TA for $J_p$>1 TW/cm$^2$ varied somewhat when the excitation conditions and beam geometries were changed between the runs (see several series of the TA measurements plotted together in Fig. 4S). As shown below, in this terawatt regime, the heat release and thermal lensing by the photoexcited sample are significant and cause irreproducibility.

### *3.2. Thermalization dynamics and the electron spectrum in the terawatt regime.*

As mentioned in the Introduction, one of the most striking kinetic transformations that occur in the terawatt regime is the change of short-term electron dynamics ($t$<10 ps) occurring during the thermalization, relaxation, and hydration of the electron. The typical evolution of $\lambda = 1.2$ μm kinetics as a function of pump radiance is shown in Fig. 3(a) (on a double logarithmic scale) and Fig. 3(b) (on a linear scale, after normalization at $t$=5 ps). As the pump radiance increases from 0.4 to 2.1 TW/cm$^2$ the relative amplitude of the initial "spike" decreases almost 3 times. This "spike" is from a short-lived precursor of the hydrated electron whose absorption spectrum in the near IR undergoes a rapid blue shift in the first 2 ps after photogeneration. [11-15,21] In Fig. 4, the ratio of the $t$=10 ps and the "spike" absorbance signals for $\lambda$=1.2 μm are plotted vs. the 400 nm light radiance. For the pump radiance < 1 TW/cm$^2$, the ratio of these two absorbance signals changes little as a function of the laser power. However, above 1 TW/cm$^2$, there is a rapid increase in the ratio with the laser power, indicating considerable flattening of the thermalization kinetics in the terawatt regime (which is also apparent from Fig. 3(b)). The remarkable feature of Fig. 4 is the abruptness of the transition between the low-radiance and high-radiance regimes.

The relative decrease in the amplitude of the "spike" with increasing pump power becomes lower at shorter wavelength of the probe light. For $\lambda$=1.05 and $\lambda$=1.1 μm kinetics, the effect is smaller (Fig. 5S); for $\lambda$<800 nm, no change of the thermalization kinetics with the pump radiance is observed (Fig. 6S). While the thermalization stage is over in 5 ps, the electron spectra obtained in the terawatt regime ($J_p$=1.9 TW/cm$^2$) at $t$=50 ps are still red-shifted with respect to the electron spectra of hydrated electrons in the room temperature water (Fig. 5(a)). As known from nanosecond pulse radiolysis studies, [16,22,23] the latter spectra undergo systematic red shift with increasing water temperature (see the Supplement for the explicit temperature dependencies of the spectral parameters). From these data, it is possible to estimate the solvent temperature corresponding to the red shift observed in Fig. 5(a). The surprising answer is that this temperature jump is very substantial, since the position of the maximum of the absorption band corresponds to a bulk temperature of 80 °C. The origin of this temperature increase and the red shift of the "thermalized" spectrum (Fig. 5(b)) is discussed in the next section. Below, we give a qualitative explanation of how the temperature jump accounts for the observed changes in the thermalization kinetics as observed by pump-probe TA spectroscopy:

When the localized, pre-thermalized electron is generated by photoionization of water, it has > 1 eV excess energy and consequently its spectrum is not too sensitive to the temperature of the medium. From the standpoint of laser spectroscopy, the thermalization process can be viewed as a systematic blue shift of the absorption band of this electron in the visible and near IR. [11-15,21] Unlike the spectrum of the initial, pre-thermalized species, the spectrum of thermalized electron strongly depends on the water temperature, systematically shifting to the red with increasing temperature. [16,22,23] Thus, in hot water the blue shift of the electron band during the thermalization/hydration process becomes progressively smaller with increasing temperature. Since the "spike" in the TA kinetics observed to the red of 800 nm is due to this spectral shift, this "spike" (after normalization by the $t$=5 ps absorbance) becomes relatively smaller in amplitude as the temperature increases. This point is exemplified by Fig. 6 in which thermalization dynamics of electrons observed at $\lambda$=1 µm and normalized at $t$=5 ps are given as function of water temperature: 28, 57, and 85 °C (for 3-photon ionization of water with 0.2 TW/cm$^2$ 400 nm light). Similar changes in the near-IR kinetics as a function of water temperature have been observed by Unterreiner and coworkers. [14] These changes look remarkably similar to those observed for room temperature water in the terawatt regime. The fact that the time profiles of thermalization kinetics observed for $\lambda$<800 nm (Fig. 6S) do not change with the laser power does not contradict this picture: First, these kinetics change little in hot water (see ref. 14). Second, our recent studies of electron thermalization in 200 nm biphotonic ionization of water suggest that the time profile of the kinetics observed at $\lambda$=450-700 nm changes little with the wavelength $\lambda$. [21] Due to this weak dependence of the time profile on the water temperature and the wavelength, a red shift of a few tens of nanometers is inconsequential.

*3.3. The temperature jump.*

In this section, the origin of the temperature jump observed in the terawatt regime is discussed. We emphasize that this temperature jump is not related to the "local" increase in temperature during electron thermalization. The latter has been postulated by Keiding and co-workers [24,25] to account for sub-picosecond relaxation of the electron, by rapid heat transfer to its surroundings. As seen from the previous section, in the terawatt regime, the water stays hot well after this relaxation stage is completed and the electron is in thermal equilibrium with its environment. In other words, the temperature jump observed in this regime is a bulk, macroscopic effect.

We suggest that the temperature jump is a natural consequence of the photophysics of 400 nm ionization. To demonstrate this point, we consider a particular set of excitation conditions given in the caption to Fig. 2. In Fig. 2(a), trace (i), the concentration profile of electrons simulated using $\phi_n$ and $\beta_n$ obtained in section 3.1. is plotted vs. the penetration depth $z$ of 400 nm, 1.6 TW/cm$^2$ light. At the surface of the jet, the electron concentration is ca. 90 mM, while the concentration averaged over the 90 μm thick jet is ca. 6 times lower. The deposition of the light energy into water follows the same profile, i.e., the concentration of electrons is maximum exactly where most energy is deposited. If all of this absorbed energy were used to ionize the solvent, relatively little heat from the electron thermalization would be deposited into water (which would account, by a conservative estimate, for the release of 1 eV of heat per ionization event). However, as shown in section 3.1, the quantum yield $\phi_n$ of ionization is only 0.42. Assuming that water molecules that do not ionize dump the excess energy into the bath, we obtain that 5.5 to 6.5 eV of the total absorbed energy (equal to 9.3 eV) ends up as heat (the exact estimate depends on how the energy released in the electron thermalization and H-OH dissociation is counted). At the surface, the density of the energy absorbed by water is ca. 190 J/cm$^3$ which is equivalent to 46 cal/g. The heat capacity $C_w$ of water is 1 cal g$^{-1}$ K$^{-1}$ and the density $\rho_w$ is 1 g/cm$^3$.[26] Provided that 60% of the absorbed energy is released as heat, the temperature of water would increase by 28 °C. Using the spectroscopic data of Bartels and coworkers,[22] it can be shown that to double the 1.1 μm absorbance of the hydrated electron, by a temperature-induced red shift of its absorption spectrum, a thermal jump of 33 °C would be sufficient. The temperature jump rapidly decreases as a function of the penetration depth of the 400 nm light. However, since the electron concentration tracks the same profile, the resulting spectral shift is still substantial. Using the parameterization of $e_{aq}^-$ spectrum as a function of temperature given by Bartels and coworkers [22] and averaging along the path of the beam, a sample-average spectrum in Fig. 5(b) was obtained. This spectrum is notably red shifted with respect to the equilibrium $e_{aq}^-$ spectrum at 25 °C.

Importantly, once the heat is released, the equilibration of the solvent across the jet would take a long time. The heat conductivity $\kappa_w$ of water is 6 mW cm$^{-1}$ K$^{-1}$ so that the heat diffusivity $\mu_w = \kappa_w/\rho_w C_w$ is 1.4x10$^{-3}$ cm$^2$/s. [26] A temperature gradient over 10 μm (the same as the concentration gradient shown in Fig. 2(a)) would take ca. 0.3 ms to disappear by diffusive heat exchange. Thus, the temperature stays constant on the sub-nanosecond time scale of our kinetic observations. Another important point is that homogenization of the heat on the microscopic scale would be rapid. In the model discussed above, the heat release and water ionization occur in different excitation events, i.e., to observe the effect of the heat (the temperature jump) on the electron spectrum and geminate recombination and thermalization dynamics, the heat released in an excitation event should reach the environment of the electron generated in a separate ionization event. It is the latter consideration that explains why the kinetic transformations observed in the terawatt regime are relatively abrupt: not only should a sufficient amount of heat be released to generate the temperature jump, the density of the excitation events should be sufficiently high as well - otherwise, the heat release in spatially isolated events would have little effect on the short-term dynamics. Again, the photophysics of tri- 400 nm photon excitation are such that the critical density of these events is achieved in the same regime where the temperature jump is substantial. E.g., the electron concentration at the surface of the jet in Fig. 2(a), trace (i), is so high that the average distance between the excitation events is ca. 1.6 nm; on this scale, thermal equilibration of the solvent would take only 3 ps.

In the estimates for the instantaneous temperature jump given above, H-OH bond dissociation of photoexcited water molecules was neglected. The branching ratio between water ionization and bond dissociation in three 400 nm photon excitation of water is not known. In bi- 266 nm photonic excitation of water, Thomsen et al. [25] estimated that the branching ratio between the dissociation and ionization is 0.55:1. Note that rapid recombination of caged H and OH radicals deposits the heat back into the solvent.

### *3.4. Geminate kinetics and the escape probability.*

Since the amplitude of the TA signal from the electron changes very rapidly with 400 nm light radiance, it is difficult to obtain good quality kinetics in the low and high power regimes using the same sample. At low power, the absorption is weak and a relatively thick sample is needed to obtain a TA signal of $(1-10) \times 10^{-3}$ OD to sample the geminate kinetics with a reasonable signal-to-noise ratio. Conversely, at intermediate and high pump power, only short-path high-speed jets could be used to avoid damage to the sample.

In the terawatt regime, the yield of the electrons is so high that cross recombination in the bulk solvent begins to occur on the sub-nanosecond time scale. This effect is mainly the consequence of inhomogeneous absorption of light. [9,17] In Fig. 2(b) *(solid lines)*, the decay kinetics for homogeneous recombination of electrons averaged along the path of the probe light are simulated for the initial concentration profiles shown in Fig. 2(a), as a function of the pump radiance, assuming a (bulk) recombination constant of $3 \times 10^{10}$ M$^{-1}$ s$^{-1}$ (close to the rate constant of $e^-_{aq}$ + $OH_{aq}$ recombination at 25 °C). A slightly greater rate constant of $3.8 \times 10^{10}$ M$^{-1}$ s$^{-1}$ for cross recombination of $e^-_{aq}$ in water ionized by the intense 266 nm laser was obtained by Pommeret et al. [17b]. For comparison, dashed lines in Fig. 2(b) give the recombination kinetics obtained assuming homogeneous generation of the electrons along the path of 400 nm light (the initial concentration had the same value as the mean concentration for the corresponding $C(z)$ profile shown in Fig. 2(a)). It is seen that the recombination kinetics are much faster for the inhomogeneous distribution, in agreement with the experimental results discussed below. Thus, at the higher end of the pump radiance, the kinetic profiles observed within the first 600 ps after the ionization are strongly affected by the cross recombination; the latter cannot be fully disentangled from the geminate recombination which occurs on the same time scale.

Fig. 7, trace (i), shows typical pump-probe kinetics for 800 nm absorbance from water excited by 400 nm light (0.1 TW/cm$^2$) obtained using a 5 mm optical path flow cell ("normal" beam geometry with $\rho_p$=92 μm and $\rho_{pr}$=12.2 μm). The initial "spike", which has the same time profile as the response function of our setup, is due to nonlinear

absorbance (simultaneous absorbance of 400 nm and 800 nm photons). As the pump radiance increases and more 400 nm light is absorbed by the sample, fewer 400 nm photons are available for this nonlinear process, and the relative amplitude of the "spike" decreases dramatically. The slow decay on time scales of 10 to 600 ps is due to the geminate recombination of hydrated electrons; heat release and cross recombination are negligible in this low-power regime. Very similar kinetics (save for the smaller amplitude of the "spike") were obtained for 0.2 TW/cm$^2$ excitation of water in a 560 μm thick high-speed liquid jet (see trace (ii) in Fig. 7; $\rho_p$=85 μm and $\rho_{pr}$=10 μm).

Figs. 8(a) and 8(b) show TA kinetic traces ($\lambda$=800 nm) obtained at higher pump radiances. Fig. 9 shows the TA signals at $t$=500 ps and $t$=10 ps plotted as a function of $J_p$ (these data were obtained using the "reverse" beam geometry: $\rho_p$=85 mm and $\rho_{pr}$=190 mm). The ratio of these two TA signals plotted in the same figure is a measure of the fraction of electrons that escape recombination at $t$=500 ps. As seen from Fig. 9, for $J_p$ > 0.3 TW/cm$^2$ this ratio first increases with the pump radiance; then, at around 1 TW/cm$^2$, it begins to decrease. The latter changeover occurs in the same range where rapid cross recombination is expected to occur, and the kinetics shown in Fig. 8 (see also Fig. 3(a), trace (iii) and Fig. 7S) reveal the long-term effects of such a recombination, which are similar to those simulated in Fig. 2(b). (Note that for the TA kinetics obtained using the "reverse" beam geometry, cross recombination is less prominent since the average concentration of electrons across the pump beam (which is probed using this geometry) is substantially lower than that at $r$=0, which is probed using the "normal" beam geometry). Even those TA kinetics that are affected by the homogeneous recombination on a sub-nanosecond time scale show the same short-term behavior as the kinetics obtained in the 0.3-1 TW/cm$^2$ range (Fig. 8(b) and Fig. 7S). In particular, within the first 100 ps after the 400 nm photoexcitation, these kinetics are much flatter than the kinetics shown in Fig. 7. To some degree, this flattening of the kinetic traces is due to the changes in the thermalization dynamics. The latter can be eliminated by comparing TA kinetics normalized at $t$=10 ps, when the (power-dependent) electron thermalization is complete. The geminate electron decay certainly slows down (Figs. 8 and 7S) and the survival probability increases (Fig. 9) with increasing radiance of the 400 nm light. Most of this change occurs in a narrow range of pump radiance, between 0.5 and 1 TW/cm$^2$. At higher radiance, this trend is difficult to follow because the cross recombination interferes with the geminate recombination. Importantly, the power range in which the geminate kinetics change with increasing $J_p$ is the same range in which the thermalization kinetics also begin to change. The latter kinetics are, actually, less sensitive to the photogenerated release of heat because a high critical concentration of the excitation events (achieved only for $J_p > 1$ TW/cm$^2$) is needed for rapid homogenization of the heat on the picosecond time scale, as noted in section 3.2.

We believe that the geminate recombination kinetics observed in the terawatt regime are for the hydrated electron in *hot* water and explain the observed transformations by the temperature jump:

The effect of increasing temperature on the geminate recombination in water has been studied by Madsen et al. [24] (who followed the electron decay kinetics to $t=90$ ps). The main effect is to increase the survival probability of the electron, from ca. 0.55 at 10 $^o$C to 0.75 at 80 $^o$C (at $t=90$ ps, see Fig. 4 in ref. 24). Interestingly, while diffusion is faster in hot water, the geminate recombination after the first 10 ps becomes slower. Madsen et al. explain these observations by the decrease in the dielectric constant of water and the fact that recombination of the electron, $e^-_{aq}$, with the OH radical is not diffusion controlled at the higher temperature, so that the probability that a diffusion encounter leads to recombination of the $(OH, e^-_{aq})$ pair decreases. Such pairs, by the estimate of Keiding and co-workers, [24,25] account for 80% of all recombination events after water photoionization; the remaining 20% are from the recombination of $e^-_{aq}$ with the hydronium ion, $H_3O^+$. Recently, Crowell et al. [18] studied the temperature effect on the dynamics of $(OH, e^-_{aq})$ pairs generated by 200 nm light induced electron photodetachment from aqueous hydroxide. The decay kinetics of the electron were followed out to 600 ps, and the temperature range of 8 to 90 $^o$C was explored (see Figs. 3 and 4 in ref. 18). The changes of the long-term kinetics with increasing temperature observed for this closely related photosystem are remarkably similar to those observed by Madsen et al. in water: [24] the electron escape becomes more efficient and the decay kinetics slow down, so that the diffusional "tail" of the kinetics becomes flatter. These trends are also present in the kinetics observed in the terawatt regime for 400 nm photoionization of neat water. While we do not have sufficient data to demonstrate by rigorous simulation that the heat release alone causes *all* of the observed kinetic transformations, [27] the qualitative behavior is consistent with the notion that the changes in the geminate kinetics are of the same origin as the concurrent changes in the thermalization dynamics.

*3.5. Implications for pulse radiolysis of water.*

In pulse radiolysis of condensed-matter systems, the excitation energy of the ionizing particle is deposited very unevenly: the excitation/ionization events are clustered in a small volume of 2-5 nm diameter (the so-called radiolytic spur); these spurs are separated by vast distances (ca. 100 nm along the particle track). [28] It has been speculated by Eisenthal and coworkers [29] and others [30] that rapid heat deposition in spurs could change the kinetics observed on the picosecond time scale, before the heat dumped into the media during electron thermalization escapes from the spur. Until recently, there was little support for such an assertion. However, recent observations from two laboratories [31,32] seem to provide such support. It appears that the temperature jump observed in multiphoton ionization of water (see above) also occurs in water radiolysis, though the dynamics of the two effects are quite different.

Recently, Gauduel et al. [31] studied the electron dynamics in room-temperature liquid $H_2O$ using a novel technique of femtosecond pump-probe pulse radiolysis - TA spectroscopy. The 700 fs FWHM pulse of relativistic (2-15 MeV) electrons was generated by interaction of a 10 TW amplified Ti:sapphire laser pulse with a supersonic helium jet. [33] The TA signal from hydrated electron was observed at 820 nm and followed over the first 150 ps after the radiation-induced water ionization. These kinetics were qualitatively different from the ones extrapolated from the pump-probe data obtained on a longer time scale, both in pulse radiolysis [34] and multiphoton excitation. [1,2,3,9,21,24,25] Specifically, instead of smooth, sloping decay curves similar to those shown in Fig. 8, a fast component with an exponential life time of ca. 20 ps was observed. Since the kinetics observed for $t > 50$ ps [29,32,34] are indicative of a broad initial distribution of electron-hole distances, the 20 ps component could not be accounted for by the geminate recombination. On the other hand, the relatively long lifetime of this component (as compared to the characteristic time of the solvation dynamics, < 300 fs) [12,13,14] excludes the involvement of pre-solvated electron. The same 20 ps component, albeit smaller in the amplitude, was observed by Wishart et al. [32] at the probe wavelength of 1000 nm using the picosecond laser cathode driven linac facility (LEAF) [35] at the Brookhaven National Laboratory (Fig. 8S(a)).

This short-term behavior might be qualitatively accounted for by a temperature jump in the radiolytic spur. [29,30] The difference between the photolysis and radiolysis is that for the latter, heating of water is localized on the nanoscale.

In the near IR, the molar absorptivity of the electron plotted vs. the local temperature passes through a maximum at > 100 $^o$C (Fig. 8S(b)). When the local temperature in the spur decreases on the picosecond time scale, this causes the *decrease* in the TA in this spectral region. Assuming that the distribution of the local temperatures follows that of the electrons (since the heat is deposited during the electron thermalization), the characteristic cooling time of a spur of the size $a$ is $a^2/\mu_w$. [30] For $a=2$ nm, an estimate of 20-30 ps is obtained - in reasonable agreement with the observed time constant for the fast component. [31,32] The substantial temperature jump is reasonable since the average energy required to ionize water by 2-15 MeV electrons (20-25 eV) [28] is more than twice the optical gap of this liquid (10-11 eV). [3,4] This excess energy is released as heat during the electron thermalization, and the local temperature immediately after the electron hydration increases substantially. Simple estimates suggest a temperature jump of at least 30-40 $^o$C at the center of the spur. [29] As seen from trace (ii), Fig. 8S(b), such a temperature jump would be sufficient to account for the fast component observed at 1000 nm in the experiments of Wishart et al. [32] A simple way to verify this explanation would be to observe the electron dynamics at other probe wavelengths.

## 4. Conclusion.

It is suggested that previously observed transformations of geminate recombination and electron thermalization kinetics in multiphoton ionization of liquid water by 400 nm terawatt light are caused by rapid heat deposition into the bulk sample. The release of heat during the photoexcitation causes a nearly instantaneous temperature jump of several tens of $^{o}$C. The ultimate cause of this jump is the low quantum yield for the 3-photon ionization of water: only 42% of the excitation events result in solvent ionization; the rest of the energy absorbed by water is released as heat. Consequently, the electron spectrum shifts to the red, the thermalization dynamics of the electron observed by TA in the near IR become flatter, the geminate recombination dynamics become slower, and the survival probability (i.e., the escape fraction) of the electrons increases.

Our analyses suggest that the ionization of water by 400 nm light remains three photon up to the radiance at which the water begins to break down; there is little evidence for the occurrence of "3+1" photoionization in which the electron or an excited water molecule generated during the ionization/excitation process absorbs one or more 400 nm photons. Though our experiments do not exclude such a possibility entirely, this photoprocess seems to be of little importance for the excitation of water by femtosecond pulses. Furthermore, it appears that the results can be understood, albeit qualitatively, without such a postulate, by water heating alone. It is likely that other reported instances of power-dependent kinetics for hydrated electron generated by water ionization in the terawatt regime can be explained in the same way. It is also possible that the fast (ca. 20 ps) component that was recently observed in the ultrafast pulse radiolysis - transient absorption studies by Gauduel et al. [31] and Wishart et al. [32] can be accounted for by a temperature jump in the radiolytic spur, as suggested almost 40 years ago by Ingalls et al [30]

## 5. Acknowledgement.

We thank Drs. C. D. Jonah, D. M. Bartels, S. Pimblott, J. R. Miller, J. F. Wishart, and V. Malka for useful discussions. IAS thanks Drs. D. M. Bartels of NDRL and J. F. Wishart of BNL for the permission to reproduce their unpublished data. The research at the ANL was supported by the Office of Science, Division of Chemical Sciences, US-DOE under contract number W-31-109-ENG-38. SP acknowledges the support of the DGA through the contract number DSP/01-60-056.

*Supporting Information Available:* (1.) Appendix: Modeling of the photophysics for "3+1" excitation. (2.) Equilibrium spectra of hydrated electron vs. the solvent temperature. (3.) Captions to Figs. 1S to 8S; (4.) Figs. 1S to 8S. This material is contained in a single PDF file that is available free of charge via the Internet at http://pubs.acs.org.

**Figure captions.**

**Fig. 1.**

(i) Transmittance $T_p$ of the 400 nm pump light *(to the left, open squares),* and (ii) TA signal $\Delta OD_{800}$ (obtained using $\lambda$=800 nm, 60 fs fwhm probe light) from the photoelectron, $e^-_{aq}$, at the delay time of $t$=5 ps *(to the right, open circles)* in tri- 400 nm photon laser excitation of the room-temperature water flowing in a high-speed optical path $L$=70 µm liquid jet. These two simultaneously determined quantities are plotted against the (peak) photon radiance $J_p$ of the incident 400 nm light. The vertical bars indicate 95% confidence limits. The beam parameters are $\rho_p$=126 µm, $\tau_p$=120 fs, and $\rho_{pr}$=24 µm. Solid lines are quantities $T_p$ and $\Delta OD_{800}$ simulated using eqs. (6) and (7) given in section 3.1 using the optimum fit parameters $\phi_3$=0.415 for the 3-photon quantum yield of photoionization and $\beta_3$=267 cm$^3$/TW$^2$ for the absorption coefficient of water; a molar absorptivity of 18500 M$^{-1}$ cm$^{-1}$ for $e^-_{aq}$ has been assumed. The dashed line (iii) drawn through the data points in the terawatt regime indicates the region where the electron absorbance increases linearly with the light radiance (and the tentative "3+1" photoexcitation might have occurred).

**Fig. 2.**

(a) Concentration profiles $C(z)$ of the photoelectrons generated by 400 nm laser pulse along the axis of the 400 nm beam. The electron concentrations, in units of 10$^{-3}$ mol/dm$^{-3}$ (mM), at the end of a 80 µm fwhm, 330 fs fwhm, 400 nm pulse are plotted for three peak radiances $J_p$ of the 400 nm light: (i) 1.56, (ii) 1.0, and (iii) 0.5 TW/cm$^2$. *(Trace (iii) is to the right, traces (i) and (ii) are to the left).* These concentration profiles are plotted against the penetration depth $z$ of the 400 nm light for a 90 µm thick water jet. These traces were simulated using the photophysical parameters given in the caption to Fig. 1. The concentrations of the electrons at the jet surface (where $z$=0) and the mean concentrations across the jet are, respectively, (i) 90 and 15.5 mM, (ii) 23.5 and 7.5 mM, and (iii) 2.9 and 1.8 mM. The higher is the light radiance, the shorter is the effective path of the excitation light. (b) Evolution of the mean electron concentration for the same photosystem, assuming cross recombination in the water bulk with a bimolecular rate constant $k_r$ of $3 \times 10^{10}$ M$^{-1}$ s$^{-1}$ (assuming no geminate recombination). Traces (i) to (iv) correspond to pump radiances of 0.2, 0.5, 1.0 and 1.56 TW/cm$^2$, respectively. The solid lines are obtained by averaging the decay kinetics $C(z,t)$ of the electron over the corresponding concentration profile: $\langle C(z,t) \rangle = \langle C(z)/(1+ktC(z)) \rangle$. The dashed lines correspond to a hypothetical system in which the electrons are generated homogeneously across the jet, with the initial concentration equal to the mean concentration obtained by averaging over the profiles given in (a), i.e. for $\langle C(z,t) \rangle = \langle C(z) \rangle / (1 + kt \langle C(z) \rangle)$. As the radiance of 400 nm light increases, spatial inhomogeneity of the electron production has progressively greater effect on the electron decay by cross recombination.

**Fig. 3.**

Pump-probe TA kinetics of photoelectrons (observed at $\lambda=1.2$ μm) generated in three 400 nm photon excitation of water (same beam geometry as that specified in Fig. 2(a)). The 400 nm light radiance $J_p$ is (i) 0.4 TW/cm$^2$ *(open squares),* (ii) 1.23 TW/cm$^2$ *(open triangles)*, and (iii) 2.1 TW/cm$^2$ *(open circles).* In (a), the kinetic traces are given on a double logarithmic plot. Note the relative decrease in the initial "spike" with increasing pump radiance and the occurrence of second order decay in trace (iii). In (b), the kinetic traces normalized at $t=5$ ps are shown. As the pump power increases, the ratio of the $t=5$ ps signal to the maximum signal in the "spike" increases three times.

**Fig. 4.**

The power/radiance dependencies of the TA signals ($\lambda=1.2$ μm) for the photosystem shown in Fig. 3. Open squares and open circles *(to the right)* indicate the TA signals from the photoelectron at $t=5$ ps and at the "spike" maximum, respectively. (The points for $J_p>1.5$ TW/cm$^2$ are not shown). The vertical bars give 95% confidence limits for these data points. The filled circles *(to the left)* give the ratio of these two absorbance signals. The smooth line drawn through these latter points is a guide for the eye. The dashed straight lines correspond to the power-law increase in the TA signal amplitude; their slopes are close to 3.

**Fig. 5.**

(a) Trace (i): experimental TA spectrum of the photoelectron observed 50 ps after tri- 400 nm photon excitation of water in a 90 μm thick jet ($\rho_p$=73 μm, $\rho_{pr}$=22 μm, $\tau_p$=176 fs) at a pump radiance of 1.9 TW/cm$^2$. This spectrum was obtained by selecting probe pulses from the white light supercontinuum using narrowband interference filters, and focussing of these different light components at the sample varied slightly across the absorption spectrum. For the multiphoton ionization, the TA signal is very sensitive to the pump and probe beam overlap, and the amplitude of the points close to 800 nm (for which the probe beam was more focussed) is somewhat greater then that in the wings of the TA spectrum. Even with this (hard-to-avoid) distortion of the TA spectrum in mind, it is apparent that the spectrum is red-shifted relative to the absorption spectrum of thermalized electron in room temperature water (trace (ii)). Trace (iii) corresponds to the spectrum of thermalized electron in 80 $^o$C water. (b) The solid line is trace (ii) reproduced from the plot above; the dashed line is the average spectrum of (thermalized) electrons that was obtained using the concentration profile trace (i) shown in Fig. 2(a) and the $e^-_{aq}$ spectrum parameterization given in the Supplement (see section 3.3 for more detail).

**Fig. 6.**

Thermalization kinetics of the electron observed using pump-probe TA spectroscopy ($\lambda$=1 μm) for three 400 nm photon excitation of water at 28 $^o$C *(open diamonds),* 57 $^o$C *(open squares),* and 85 $^o$C *(open circles).* The peak radiance $J_p$ of the 400 nm light was fairly low, $J_p$=0.2 TW/cm$^2$, so that the light-induced temperature jump was negligible. To facilitate the comparison of the time profiles, these TA kinetics are normalized at $t$=5 ps, at which delay time the thermalization stage is complete. As the water temperature increases, the relative amplitude of the "spike" decreases.

**Fig. 7.**

Pump-probe TA kinetics of the electron generated by three 400 nm photon excitation of water in (i) a 5 mm optical path cell ($J_p$=0.125 TW/cm$^2$; *open circles, to the right*) and (ii) a 560 μm thick liquid jet ($J_p$=0.2 TW/cm$^2$; *open squares, to the left*). The solid curve is the simulation of the geminate recombination dynamics using the IRT model given by Pimblott and coworkers, ref. 36 (the model parameters are given in the caption to Fig. 1S in the Supplement; see also ref. 2) for an initial r$^2$-Gaussian electron distribution, eq. (A1), with a width parameter $\sigma_0$=1.1 nm.

**Fig. 8.**

(a) The evolution of the TA kinetics of the electron ($\lambda$=800 nm, $\rho_p$=83 μm, $\tau_p$=200 fs, $\rho_{pr}$=232 μm) generated by 400 nm photoionization of water in a 90 μm thick jet as a function of the pump radiance for $J_p$= (i) 0.29, (ii) 0.65, (iii) 1.3, and (iv) 2.6 TW/cm$^2$. Note that the "reverse" beam geometry was used for these kinetic measurements. (b) A comparison between the time profiles of normalized kinetic traces (i) *(open circles)* and (iv) *(solid line)* from Fig. 8(a).

**Fig. 9.**

Pump radiance dependencies of the TA signals at (i) $t$=10 ps *(open squares; to the left)* and (ii) $t$=500 ps *(open diamonds; to the left)* in three 400 nm photoexcitation of water in a 90 μm thick jet ($\lambda$=800 nm, $\rho_p$=85.4 μm, $\tau_p$=200 fs, $\rho_{pr}$=190 μm; note that "reverse" beam geometry was used ). Trace (iii) *(open circles; to the right)* is the ratio of the $t$=500 ps and $t$=10 ps absorbance signals. The solid line is a guide to the eye.

**References.**

\* To whom correspondence should be addressed: *Tel* 630-252-8089, *FAX* 630-2524993, *e-mail:* rob_crowell@anl.gov.

2 presently at the Experimental facility Division, Advanced Photon Source, Argonne National Laboratory, 9700 S. Cass Ave., Argonne IL 60439; *Tel* 630-252-5874, *FAX* 630-2529303, *e-mail:* jqian@aps.anl.gov.

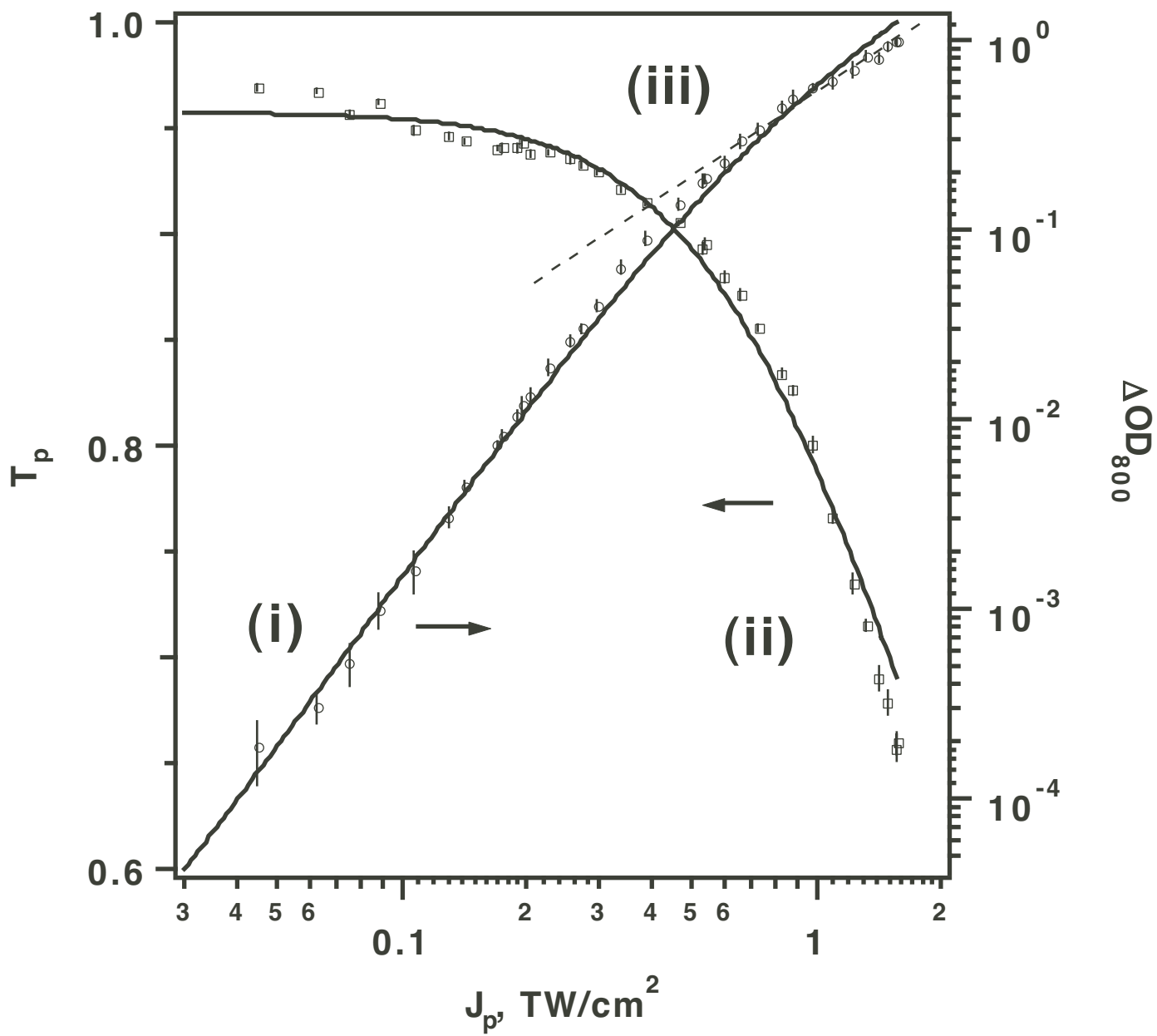

Fig. 1; Crowell et al.

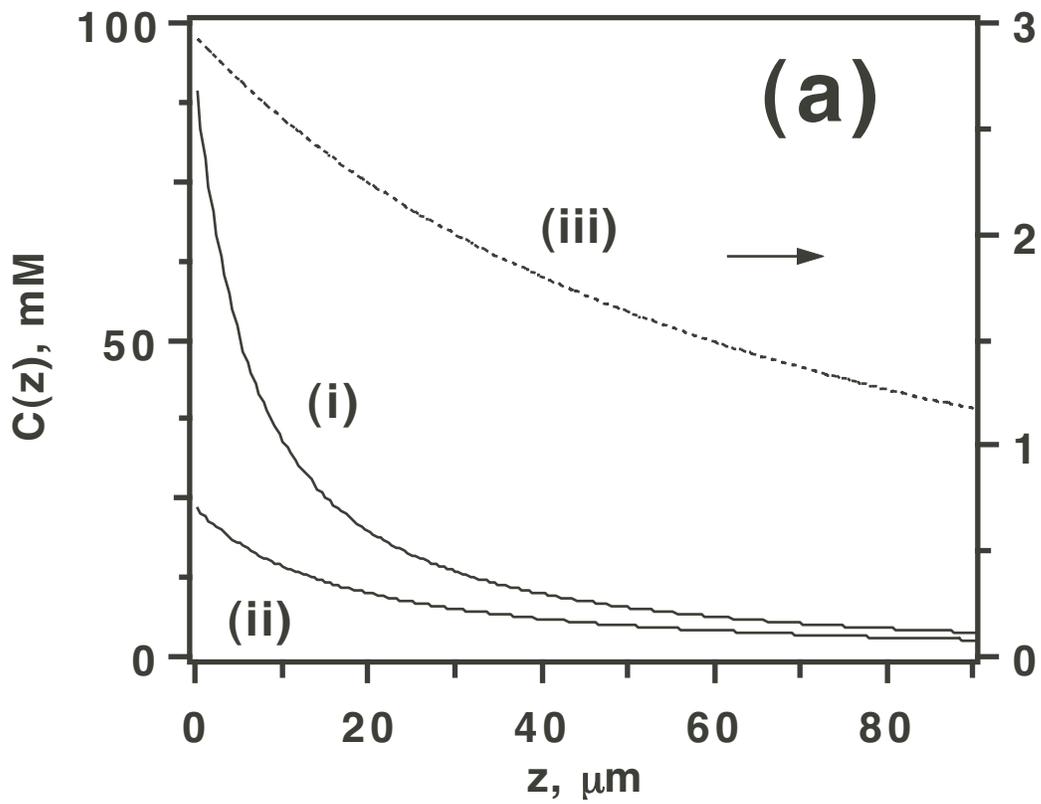
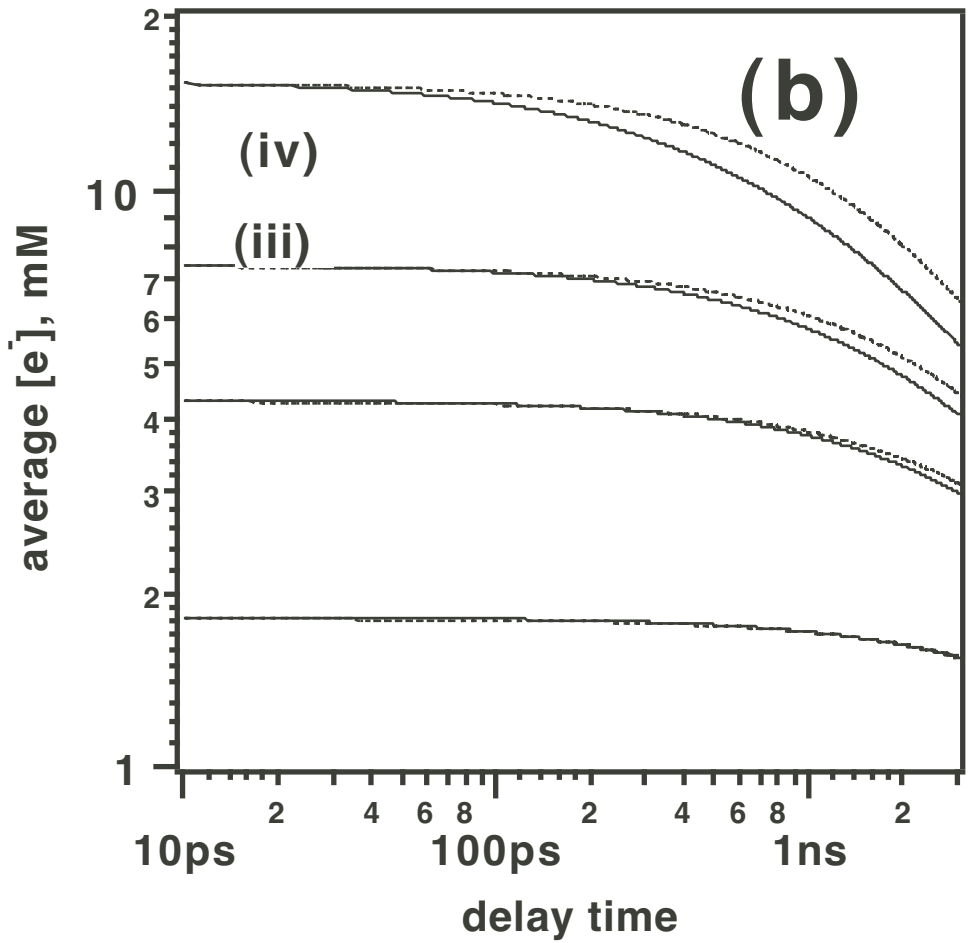

Fig. 2; Crowell et al.

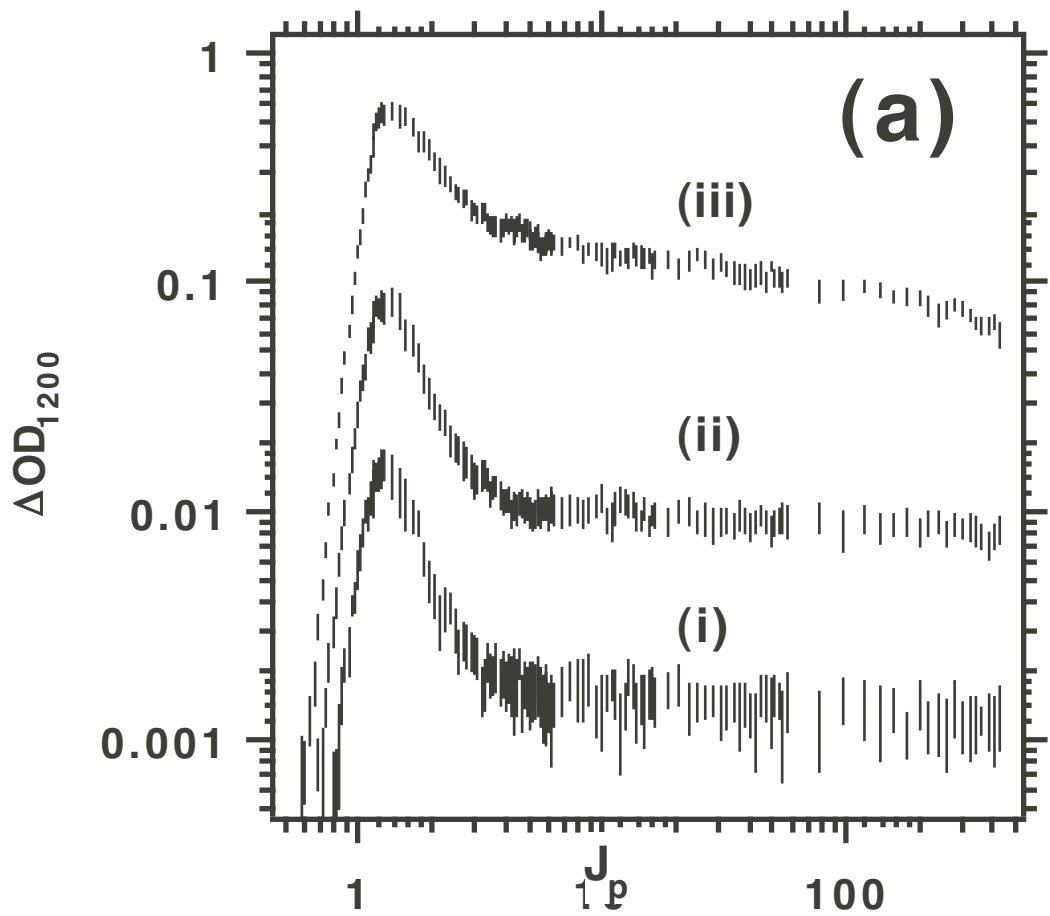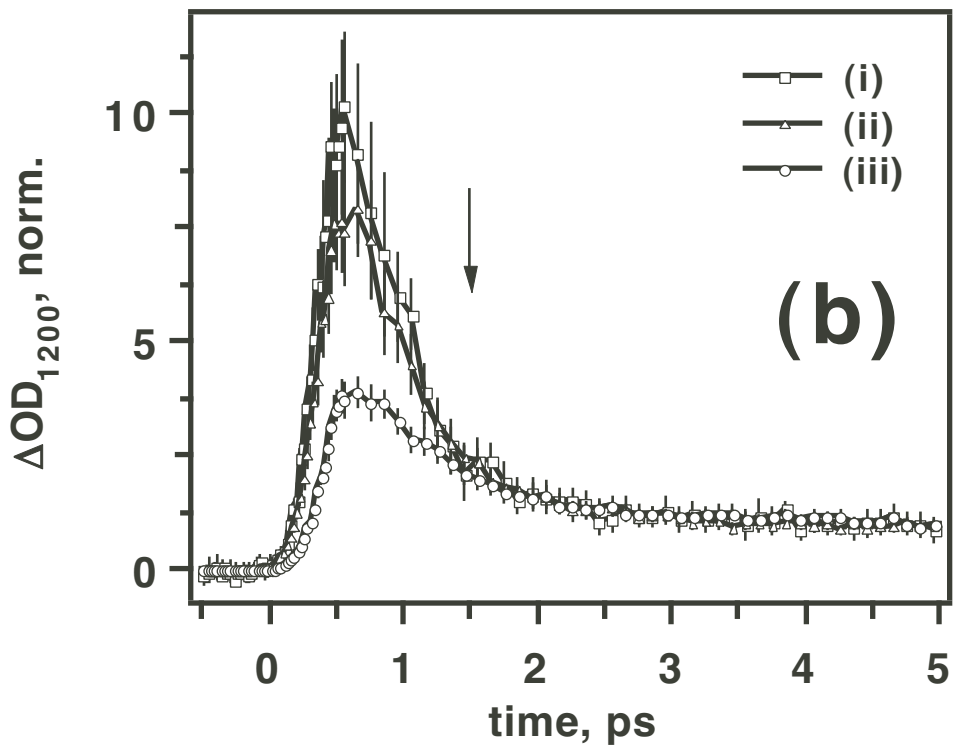

Fig. 3; Crowell et al.

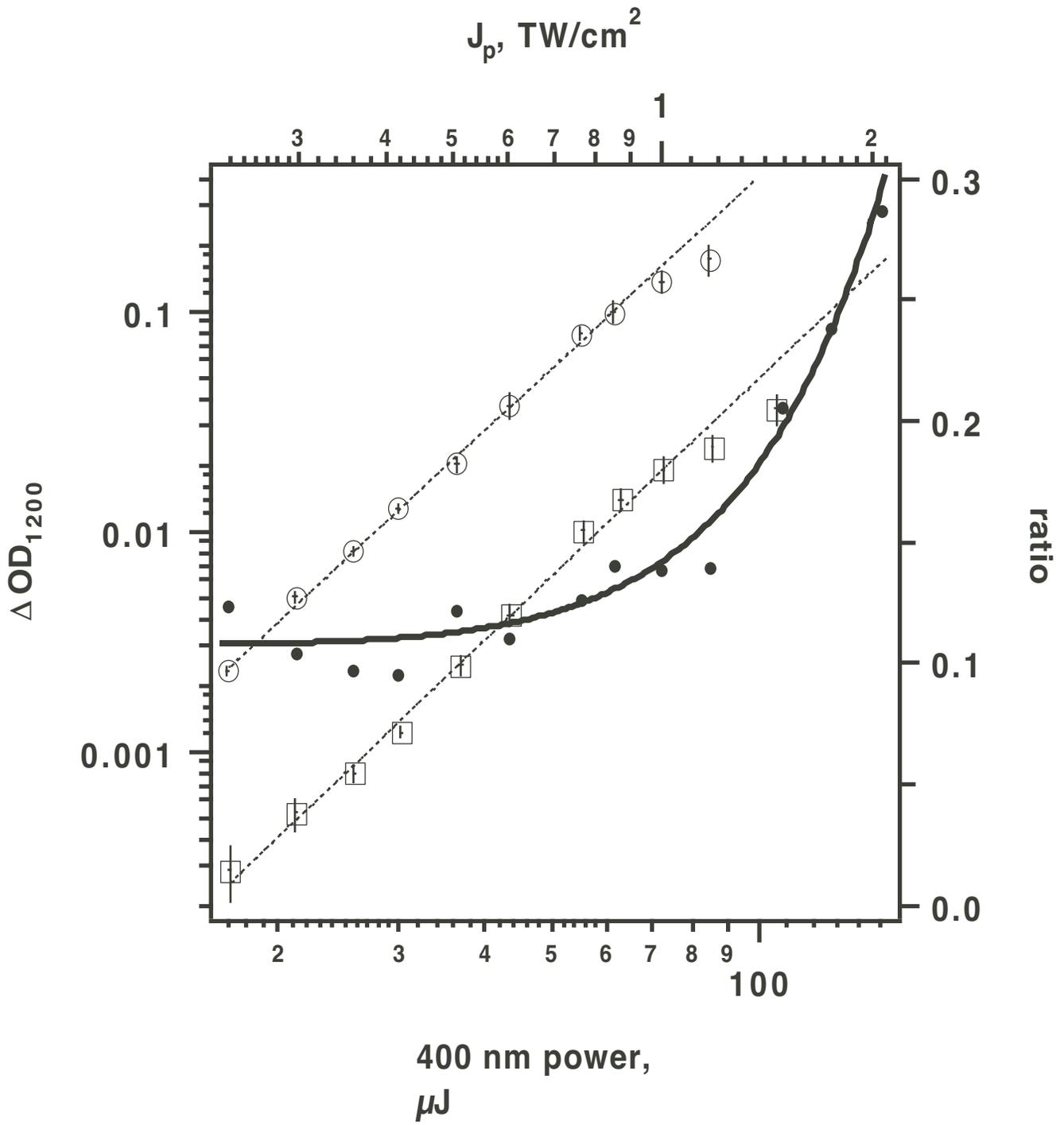

Fig. 4; Crowell et al.

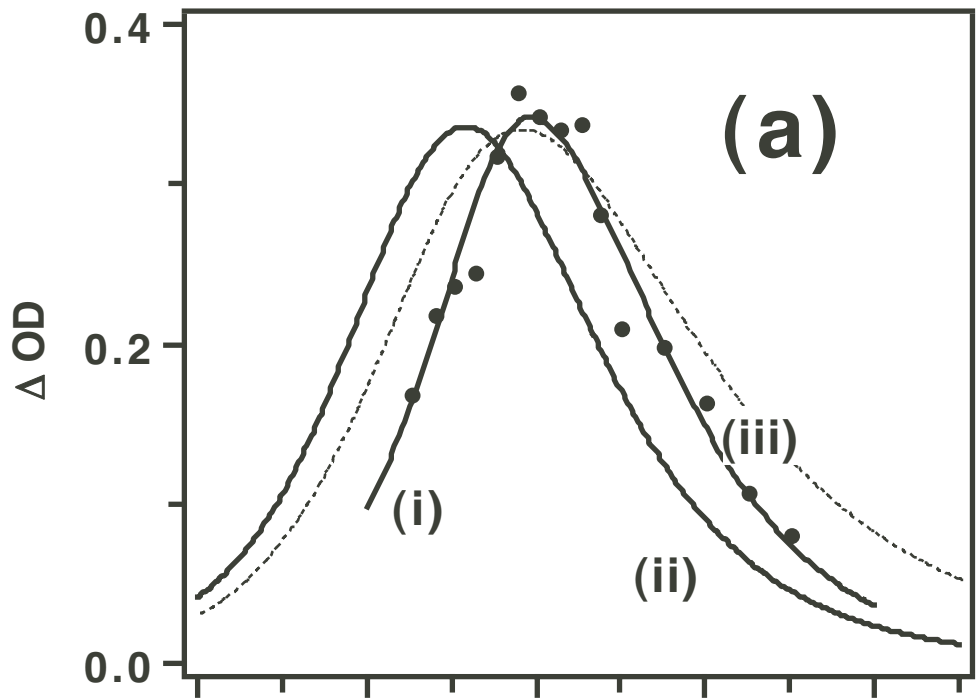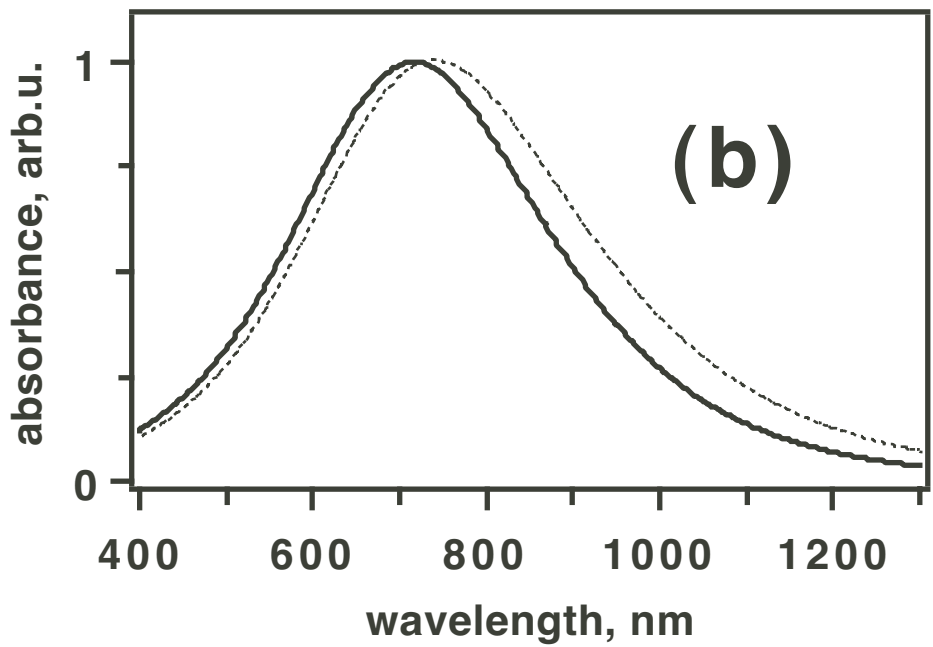

Fig. 5; Crowell et al.

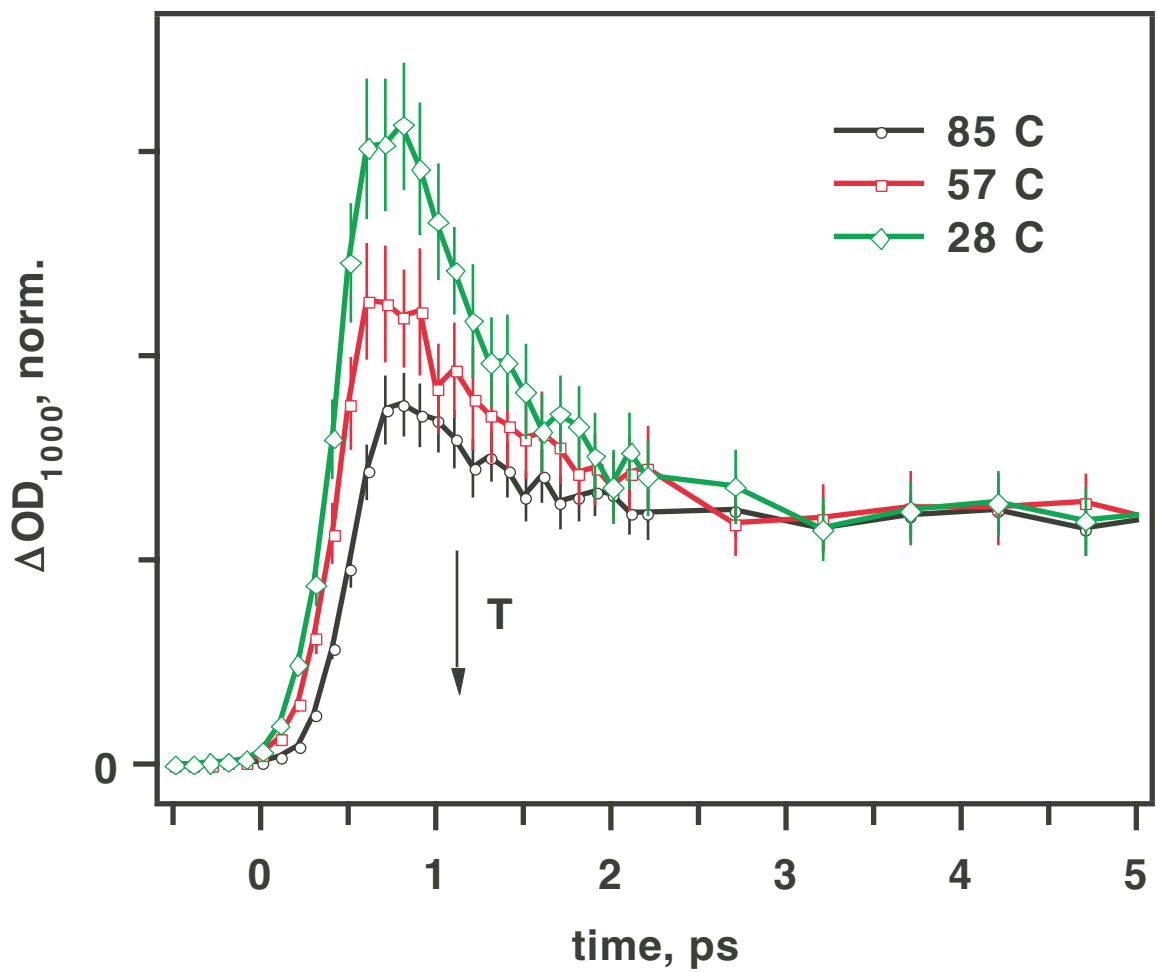

Fig. 6; Crowell et al.

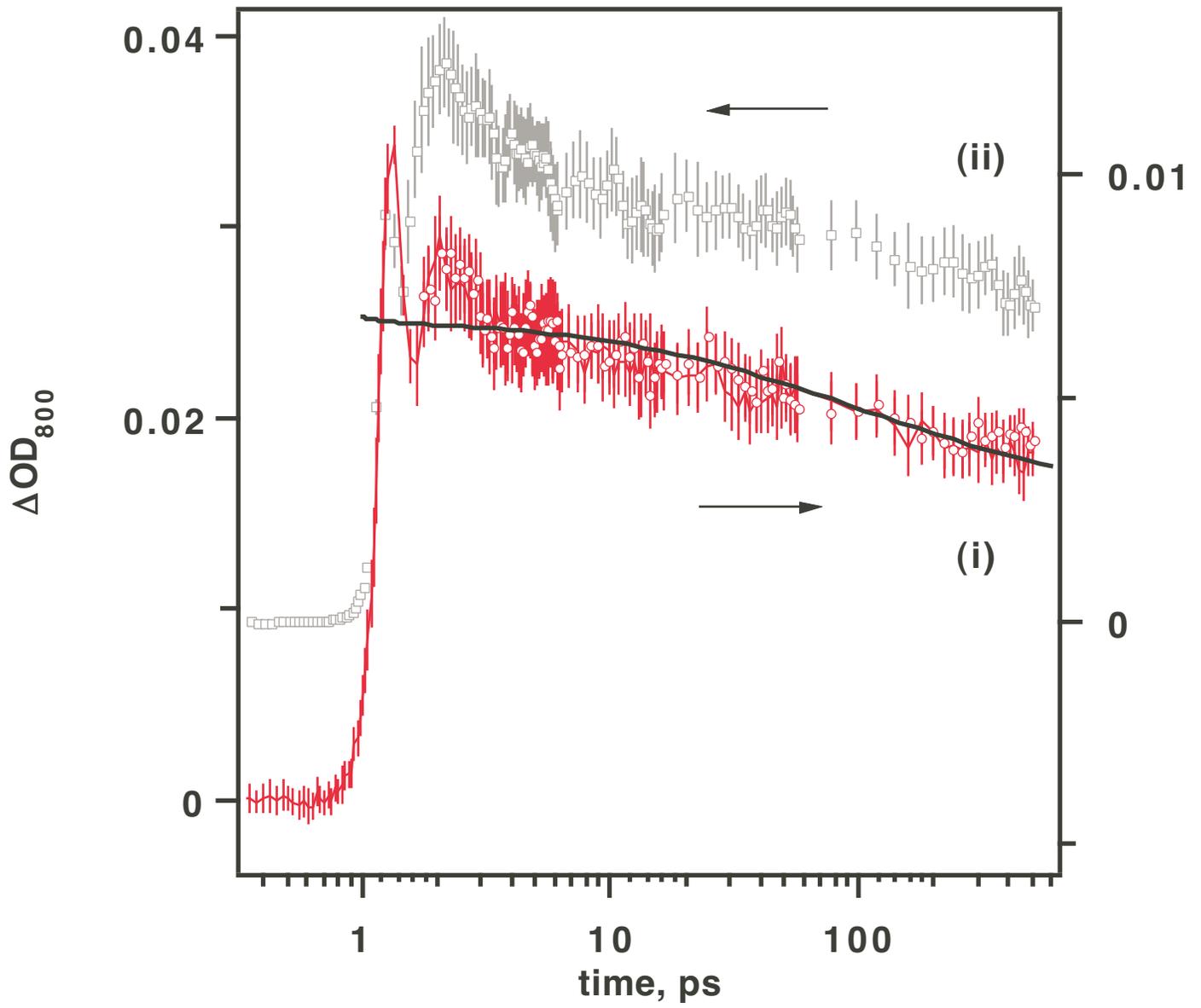

Fig. 7; Crowell et al.

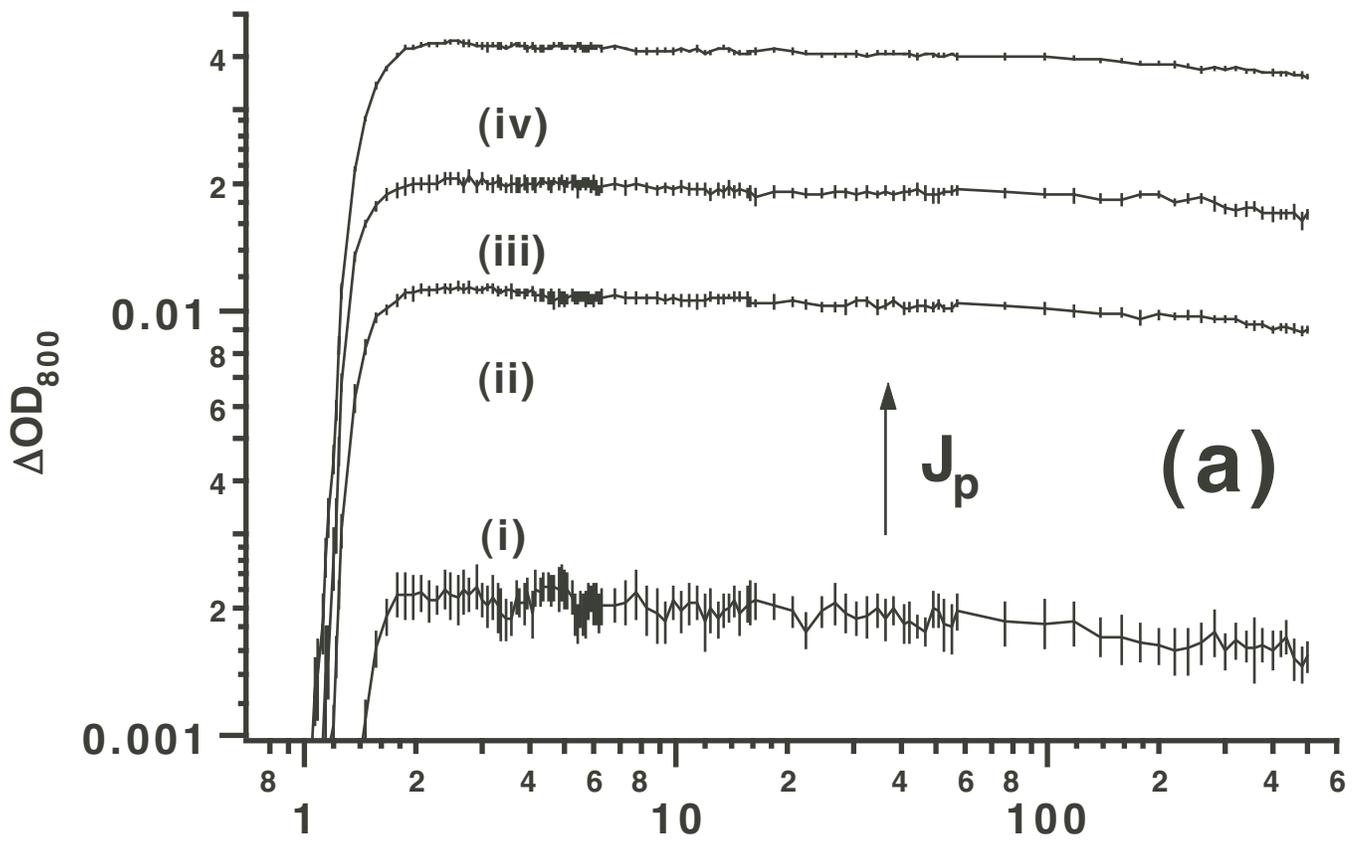
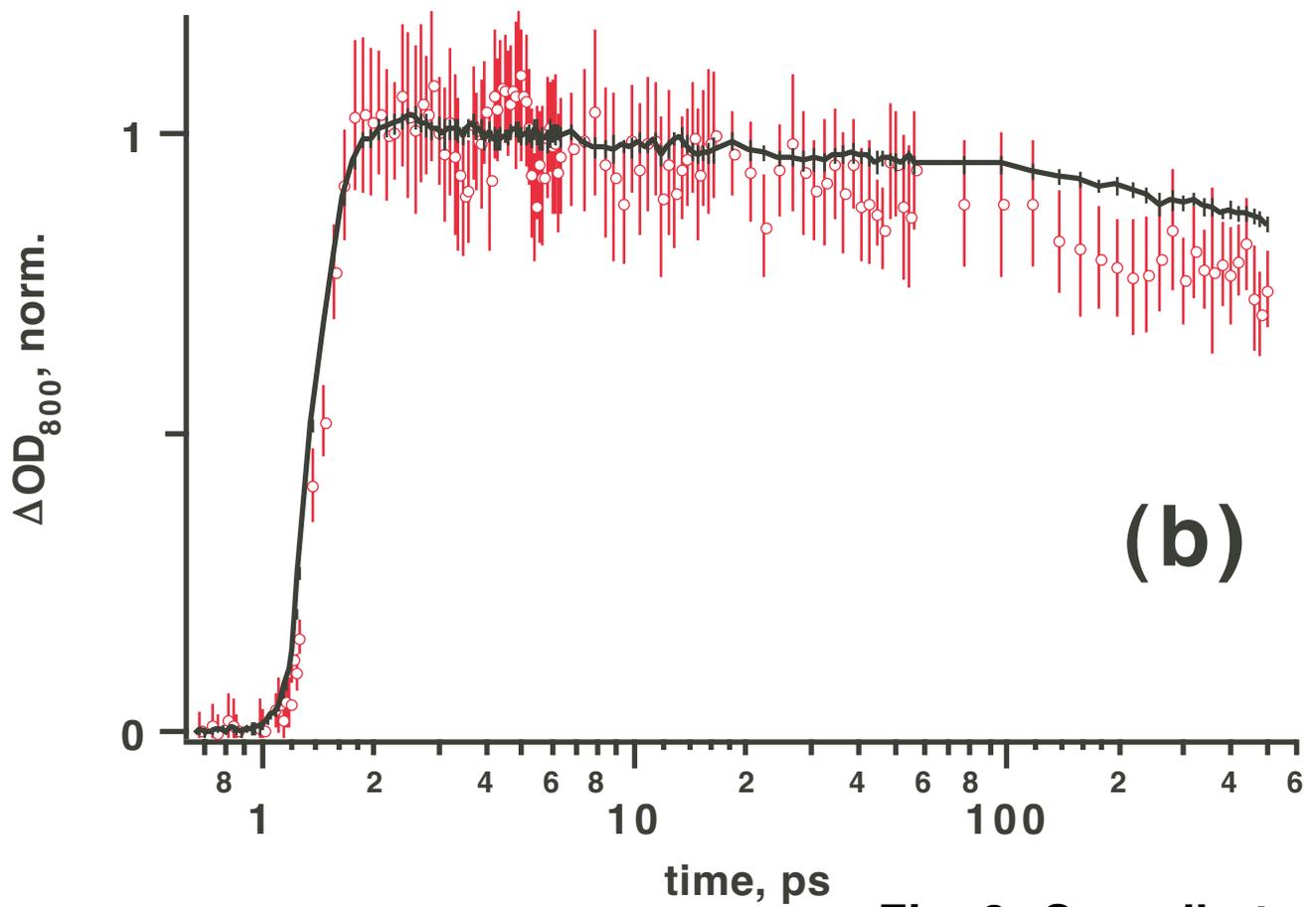

Fig. 8; Crowell et al.

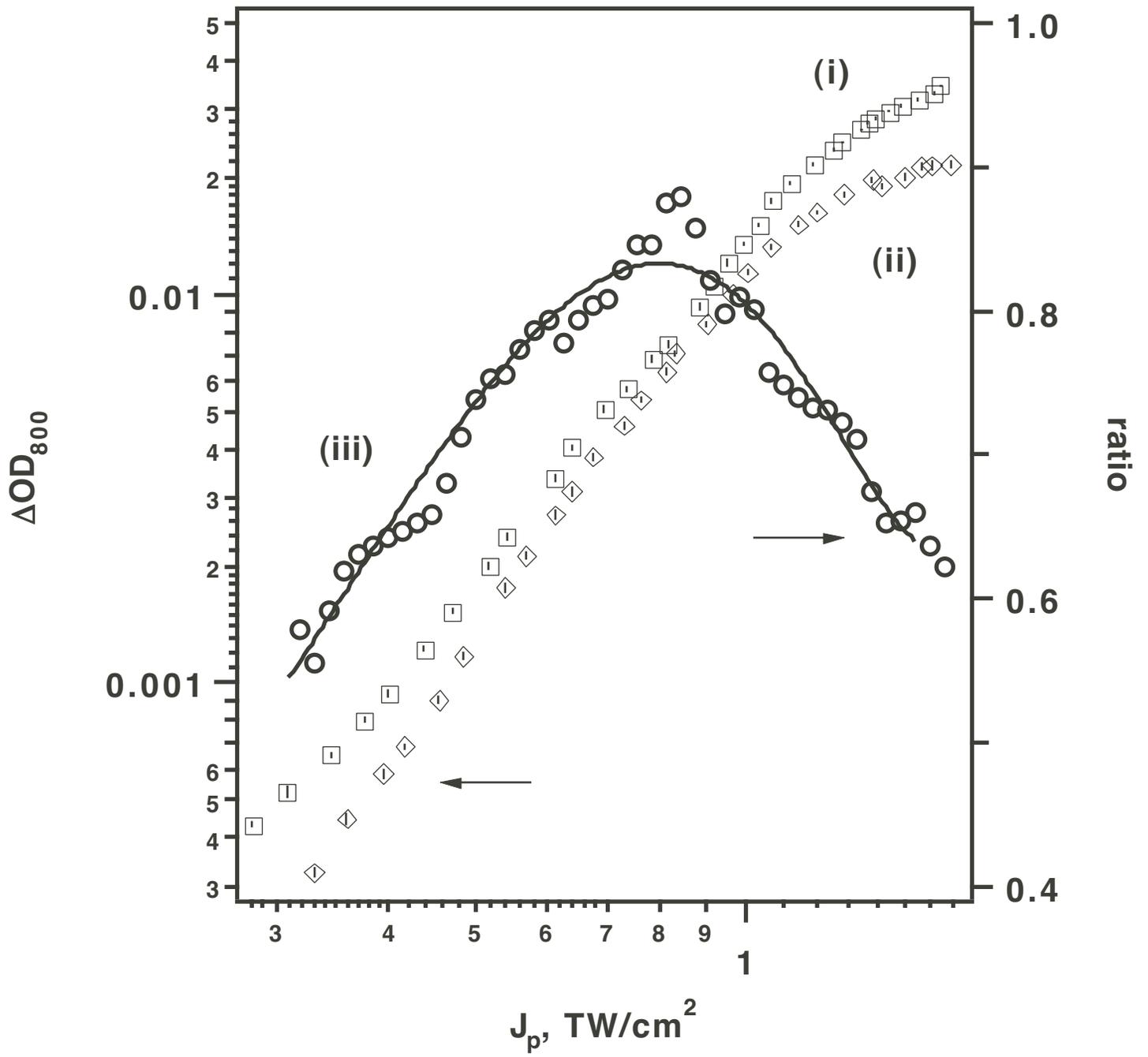

Fig. 9; Crowell et al.

**SUPPLEMENTARY MATERIAL**                          **JP0000000**

*Journal of Physical Chemistry A, Received:* *********

**Supporting Information.**

**(1.) Appendix: Modeling of the photophysics for "3+1" excitation.**

Below we consider the extension of the simple model given in section 3.1 in the text to include the possibility of a "3+1" photoprocess (i.e., the absorption of "extra" 400 nm photons by products generated in the course of photoexcitation and ionization). Though the approach given below is formulated assuming that the 400 nm light absorbing species is the electron, it is easy to use the same type of model for other species, e.g., a short-lived excited state of water.

We will assume that 3-photon ionization of water yields an electron of type $e_0^-$ which can absorb an extra 400 nm photon, thus being injected deep into the conduction band (CB). The resulting (mobile) CB electron $e_{CB}^-$ with lifetime $\tau_f \ll \tau_p$ rapidly thermalizes and localizes after migrating over the mean distance $\sigma_L$, yielding the electron $e_1^-$. The latter has exactly the same photophysical properties as $e_0^-$; the difference between these two electron states is their spatial distribution with respect to the hole. We may consider further photoexcitation of the electron $e_1^-$ into the CB; the subsequent relaxation of the CB electron would yield the electron of type $e_2^-$, etc. The distribution function of the electron before the first photoexcitation $P_0(r) = P(r, \sigma = \sigma_0)$ is $r^2$-Gaussian, where

$$4\pi r^2 P(r,\sigma) \, dr = (\pi\sigma)^{-3/2} \exp(-r^2/\sigma^2) \, d^3\mathbf{r} \qquad (A1)$$

and $\sigma_0$ is the pulse width. Since the convolution of the Gaussian functions is itself a Gaussian function, distribution functions $P_m(r)$ for the electrons of type $e_n^-$ are given by eq. (A1) with $\sigma_m^2 = \sigma_0^2 + m\sigma_L^2$. The overall electron distribution by the end of the excitation pulse is given by $P_{av}(r) = \sum p^{(m)} P(r, \sigma_m)$, where $p^{(m)}$ is the relative concentration of the electrons of type $e_m^-$. If the time and penetration depth dependence of the light intensity are neglected, it is easy to demonstrate that $p^{(m)} = p^m e^{-p}/m!$ are given by a Poisson distribution, where $p = \langle p^{(m)} \rangle = \sum_{m=1}^{\infty} m p^{(m)}$ is the mean number of 400 nm photons absorbed per electron. In Fig. 1S, geminate recombination dynamics for the electron in water are plotted as a function of this parameter. To simulate these kinetics, the independent reaction time (IRT) model of spurs in water developed by Pimblott [36] was used for the initial electron distribution given by a function $P_{av}(r)$ for $\sigma_0$=0.9 nm and $\sigma_L$=2 nm; other simulation parameters (such as the diffusion coefficients and reaction radii at 25 °C) were taken from refs. 2 and 36. As seen from this plot, the kinetics change gradually with increase in the parameter $p$, and the observed increase in the survival probability of the electron on sub-nanosecond time scales in the terawatt regime (see Figs. 8 and 7S) would require that on average at least one 400 nm photon is absorbed by the electron.

Assuming that this Poisson distribution also holds in the general case, we will neglect the subsequent excitation of the electrons and examine a two-state model that involves only electrons $e_0^-$ and $e_1^-$. To truncate the series of kinetic equations corresponding to each subsequent excitation, we will assume that the excitation of electron $e_1^-$ results in the relaxation of the CB electron to the same state $e_1^-$. Another assumption is that the CB electrons absorb neither the pump nor the probe light. The concentrations $C_m(z,r;t)$ ($m=0,1$) of electrons $e_m^-$ and the CB electron $C_{CB}(z,r;t)$ are given by the equations

$$\partial C_0/\partial t = 1/3\,\phi_3\beta_3 I^3 - \phi_1\beta_e[C_0 - C_{CB}]I, \tag{A2}$$

$$\partial C_{CB}/\partial t = +\phi_1\beta_e[C_0 + C_1 - 2C_{CB}]I - C_{CB}/\tau_f, \tag{A3}$$

$$\partial C_1/\partial t = -\phi_1\beta_e[C_1 - C_{CB}]I + C_{CB}/\tau_f, \tag{A4}$$

where $\beta_e$ is the molar extinction coefficient of electrons $e_0^-$ and $e_1^-$ at 400 nm and $\phi_1 \approx 1$ is the quantum yield for photoexcitation of these electrons into the CB. The evolution of the amplitude $I(z,r;t)$ of 400 nm light is given by

$$\partial I/\partial z = v^{-1}\,\partial I/\partial t - \beta_3 I^3 - \beta_e[C_0 + C_1 - 2C_{CB}]I, \tag{A5}$$

where $v$ is the velocity of the 400 nm light in the sample. The second and third terms in eq. (A5) correspond to the absorption of 400 nm light by water and electrons, respectively. Integrating these two terms over time and the volume of the sample yields the total energies $I_w$ and $I_{el}$ absorbed by water and electrons, respectively. The ratio $\xi_{energy} = I_{el}/(I_w + I_{el})$ gives the fraction of the total pump energy absorbed by the electrons. Similarly, one can introduce the ratio $\xi_{conc} = \langle C_1 \rangle/\langle C_0 + C_1 \rangle$ of the mean concentrations of the two kinds of the electron at $t = +\infty$. The latter ratio gives a reasonable estimate for the parameter $p$ introduced above. For obvious reasons, these two ratios closely track each other, with $\xi_{energy} \approx 1/3\,\xi_{conc}$. The absorption signal $\Delta T_{pr}$ from the electrons $e_0^-$ and $e_1^-$ is given by expression

$$-\Delta T_{pr} = \frac{2}{\rho_{pr}^2} \int_{-\infty}^{+\infty} dt \int_0^{\infty} dr\, r\, \exp\left(-\frac{r^2}{\rho_{pr}^2} - \beta_e \int_0^L dz\,[C_0 + C_1 - 2C_{CB}]\right), \tag{A6}$$

and the transmission of the 400 nm light is given by eq. (6) in section 3.1. For a given polar coordinate $r$, eqs. (A2) to (A5) were solved numerically on a lattice of 200-500 $z$-points, by stepping time $t$ in increments of 1 fs. Eqs. (A2) to (A5) can be readily generalized to include the finite lifetime of the photoproduct; however, the efficiency of the "3+1" process increases for long-lived extra-photon absorbers, and we are interested in the most favorable scenario for such a photoprocess. For the same reason, since the efficiency of the "3+1" process increases for shorter lifetime $\tau_f$ of the conduction band electron, we assumed that this lifetime is very short, ca. 50 fs. Again, this (unrealistically) short time was assumed to estimate the maximum possible efficiency of the "3+1" photoexcitation. Finally, it was assumed that pre-thermalized electrons that absorb the extra 400 nm photon absorb at this photoexcitation wavelength as much as fully thermalized electrons (with a decadic extinction coefficient of ca. 2600 M$^{-1}$ cm$^{-1}$).[16] This assumption is manifestly incorrect: according to other studies,[11-15,21] pre-thermalized electrons absorb in the blue at least 5-10 times less than the fully hydrated electrons. Once more, our goal is to assess the feasibility of the "3+1" photoprocess in the most favorable (rather than the most realistic) scenario.

Fig. 2S shows the comparison between the plots of $\Delta OD$ and $T_p$ vs. the pump radiance $J_p$ for (i) the model of section 3.1 and (ii) for the "3+1" model given above (other simulation parameters are given in the caption). Typical concentration profiles of electrons $e_0^-$ and $e_1^-$ at $r = 0$ and $t = +\infty$ for four values of $J_p$ are shown in Fig. 3S(a). It is apparent from the latter plot that efficient electron excitation in the terawatt regime occurs mainly near the water surface and requires $J_p$>0.5-1 TW/cm$^2$ to be efficient. As seen from Fig. 2S, even though we have considered the most favorable scenario for the "3+1" photoexcitation, the changes in these dependencies due to the electron excitation are very slight. The reason for that becomes clearer from Fig. 3S(b) where the ratios $\xi_{energy}$ and $\xi_{conc}$ are plotted vs. the pump radiance. Even at $J_p$=1.6 TW/cm$^2$, the fraction of 400 nm photons absorbed by the electrons is ca. 15% of the total energy. On the other hand, the conversion of $e_0^-$ into $e_1^-$ at this high radiance is substantial, ca. 50%. Our simulations suggest that no reasonable combination of photophysical parameters can give a greater conversion rate; in fact, it should be much smaller, given the unrealistic estimates used in the simulations. This means that the geminate kinetics originate from two distinct populations of the electrons: those that absorbed one or more 400 nm photons (which are always a minority) and those that did not (which constitute the largest fraction); the former type has slower decay kinetics and higher probability of escape. The resulting overall kinetics (for $p$<0.5) are not too different from the $p$=0 kinetics in the absence of electron excitation (see IRT simulation in Fig. 1S). The transition from one type of the kinetics to another is smooth because $\xi_{conc}$ is a slow function of $J_p$ (Fig. 3S(b)).

These expectations are poorly supported by our observations. The transformation of the kinetics is rather abrupt and the recombination kinetics are slower and the escape fractions are higher than simulated. The photophysics is such that the conversion of $e_0^-$ to $e_1^-$ is prohibitively low, at most 10-50%, even at the highest radiance of 400 nm photons (1-2 TW/cm$^2$) and the most generous estimates for the efficiency of the "+1" photoexcitation. We reach the conclusion that the "+1" photoexcitation of electrons generated within the duration of a femtosecond pulse is unlikely, and the kinetic behavior expected from the occurrence of such a photoprocess has not been observed experimentally. The experimental power dependencies shown in Fig. 2S can be accounted for without the tentative "3+1" photoprocess (Fig. 1), and postulating such a photoprocess does not improve the fit quality; in fact, it has little effect on these dependencies (Fig. 2S). All of these considerations are aside from the fact that the tentative "3+1" mechanism does not account for the results of section 3.2 that indicate the occurrence of a temperature jump in the terawatt regime. While we cannot completely exclude that "3+1" excitation occurs in short-pulse photoionization of water by 400 nm light, this photoprocess seems to have a relatively minor effect. An open question remains, whether the "3+1" photoexcitation plays a more prominent role when relatively long-duration (picosecond) pump pulses of high peak power are used to ionize water.

**(2.) Equilibrium spectrum of hydrated electron vs. solvent temperature.**

The parameterization of the absorption spectrum of fully thermalized, hydrated electron was taken from the unpublished work of Bartels and coworkers. The spectrum $S(E)$ of the electron in $H_2O$ as a function of the photon energy $E$ is given by a Gaussian curve on the red side and a (modified) Lorentzian curve on the blue side:

$$S(E) = S_{max} / \left(1 + \left[(E - E_{max})/W_L\right]^\nu\right) \quad for \quad E > E_{max}, \quad (A7)$$

$$S(E) = S_{max} \exp\left(-\ln 2 \left[(E - E_{max})/W_G\right]^2\right) \quad for \quad E < E_{max} \quad (A8)$$

where $S_{max}$ is the absorbance at the band maximum at $E = E_{max}$ (ca. 21000 $M^{-1}$ $cm^{-1}$ for all temperatures), $W_L$=0.51 eV is the temperature-independent Lorentzian width, and the temperature dependence of the energy $E_{max}$, the Gaussian width $W_G$, and the exponent $\nu$ are given by polynomial expansions

$$E_{max}(eV) = 1.79 - 0.0025407t + 2.5384 \times 10^{-17} t^5, \quad (A9)$$

$$W_G(eV) = 0.343 + 0.000290411t + 6.613 \times 10^{-6} t^2 \\ - 4.10413 \times 10^{-8} t^3 + 5.33721 \times 10^{-11} t^4, \quad (A10)$$

$$\nu = 1.9488 + 0.0012895t \quad (A11)$$

where $t$ is the temperature in °C. Eqs. (A7) to (A11) give the shape of the absorption band of hydrated electron with an accuracy better than 1% for temperatures between 20 and 300 °C.

**Figure captions.**

**Fig. 1S.**

A simulation of geminate recombination dynamics for hydrated electrons generated by short-pulse photoionization of 25 °C water using the IRT model of Pimblott.[30] In these simulations, the initial electron distribution is given by equation $P_{av}(r) = \sum p^{(m)} P(r, \sigma_m)$ (see eq. (A1) and the discussion in the Appendix), where $p^{(m)} = p^m e^{-p}/m!$ and the parameter $p$ is the average number of 400 nm photons absorbed by the electron in the "3+1" photoexcitation model. A family of kinetic profiles for parameter $p$ between 0 and 2.5 *(from bottom to top)* is shown. The simulation parameters are $\sigma_0$=0.9 nm [1,2,3] and $\sigma_L$=2 nm,[1,10] the average $H_3O^+ - OH$ distance of 0.28 nm,[36] the reaction radii of 0.5 nm and 0.54 nm for recombination of the electron with the hydronium ion and hydroxyl radical, respectively, and the diffusion coefficients of 4.58x10$^{-5}$, 9x10$^{-5}$, and 2.8x10$^{-5}$ cm$^2$/s for $e_{aq}^-$, $H_3O^+$, and $OH$ radical, respectively.[2] The reaction of the electron with the $OH$ radical is assumed to be diffusion-controlled at 25 °C; the reaction with the hydronium cation has a reaction velocity of 4 m/s at the recombination radius.[2,36]

**Fig. 2S**

Same as Fig. 1 in section 3.1. The dash-dot curves are simulated using the model given in section 3.1; the solid lines are simulated using the model of "3+1" photoexcitation process discussed in the Appendix. A molar absorptivity of 2610 M$^{-1}$ cm$^{-1}$ for the photoelectron and a lifetime of 50 fs for the conduction band electron were assumed in this simulation; other parameters are given in the caption to Fig. 1.

**Fig. 3S.**

(a) Simulated concentration profiles at $t = +\infty$ (i.e., at the end of the photoexcitation pulse) and $r = 0$ (i.e., the axis of the pump beam) for photoelectrons generated by $\rho_p$=80 μm, $\tau_p$=200 fs, 400 nm laser pulse vs. the penetration depth $z$ of the light. These profiles are calculated for four radiances $J_p$ of the 400 nm light (that are given in the color scale in the plot). The *(bold)* solid and *(thin)* dotted lines correspond to millimolar concentrations of electrons $e_0^-$ and $e_1^-$, respectively (see the Appendix for more details). (b) Calculated ratios $\xi_{conc}$ *(to the left)* and $\xi_{energy}$ *(to the right)* vs. the pump radiance $J_p$ obtained using the parameters of Figs. 1 and 2S and the "3+1" photoexcitation model discussed in the Appendix.

**Fig. 4S.**

Double logarithmic plot of the TA signal from the photoelectron ($t$=5 ps, $\lambda$=800 nm) in tri- 400 nm photonic ionization of liquid water in a 70 μm thick jet vs. the fluence of 400 nm photons ($\tau_p$=114 fs and $\rho_{pr}$=24 μm for all series). Several series of TA data obtained on different days using different beam geometries are plotted together to illustrate the reproducibility and robustness of such data (e.g., the pump beam radius $\rho_p$ varied from 80 to 210 μm). The straight line drawn through the data points corresponds to a slope (effective photon order) of 2.9±0.05. When the 400 nm light radiance $J_p$ exceeds 1 TW/cm$^2$, the power dependencies obtained under slightly different photoexcitation conditions diverge slightly. Nevertheless, all such power dependencies exhibit stunted growth of the TA signal with increasing radiance, as compared to the low-power dependencies. We attribute the slight irreproducibility of the TA data in the terawatt regime to the high heat load on the sample and thermal lensing.

**Fig. 5S.**

Pump-probe TA kinetics for photoelectrons generated by tri- 400 nm photon ionization of room temperature water in a 90 μm thick high-speed jet. The probe light wavelengths are (a) 1.1 μm and (b,c) 1.0 μm. These TA kinetics were obtained using a $\rho_p$=80.3 μm, $\tau_p$=200 fs pump and a $\rho_{pr}$=17.2 μm probe (i.e., "normal" beam geometry, see section 2). (a) Kinetic traces (i) and (ii) are given on different scales in order to juxtapose the maxima. Traces (i) *(squares, to the right)* and (ii) *(circles, to the left)* correspond to pump radiances of 0.35 and 1.67 TW/cm$^2$, respectively. (b,c) The time evolution of the TA signal on (b) long and (c) short time scales. In (c), the TA signal is normalized at $t$=5 ps, to facilitate the comparison between the traces obtained for different $J_p$: (i) 0.39 TW/cm$^2$ *(open diamonds)*, (ii) 1.23 TW/cm$^2$ *(open squares),* and 2.1 TW/cm$^2$ *(open circles).* Note the logarithmic time scale in (a) and (b).

**Fig. 6S.**

Normalized pump-probe kinetics of TA for $\lambda$=500 nm (a), 600 nm (b), and (c) 700 nmThe TA signal is from a photoelectron generated by 3 x 400 nm photoexcitation of liquid water in a 90 μm thick jet. These kinetics are normalized at $t$=5 ps. The 400 nm pulse is 300 fs fwhm ($\rho_{pr}$=22 μm). The filled circles and empty squares correspond to 400 nm photon fluences of (a) 1.16 and 0.18 J/cm$^2$ and (b,c) 0.9 and 0.18 J/cm$^2$, respectively. For this particular experiment, the pump beam was elliptical, with *1/e* semiaxes of (a) 76 and 71 μm and (b,c) 106 and 65 μm at the front of the jet. Observe that the time profile of the thermalization kinetics for $\lambda$<800 nm is independent of the pump radiance. Compare these kinetic traces with those for $\lambda$>800 nm shown in Figs. 3(b) and 5S.

**Fig. 7S.**

Pump-probe TA kinetics of the electron generated by 3 x 400 nm photoexcitation of heavy water, D$_2$O, in a 300 μm thick liquid jet ($\lambda$=800 nm, $\rho_p$=200 mm, $\tau_p$=150 fs, $\rho_{pr}$=50 μm). The estimated pump radiances are (i) 0.12, (ii) 0.66, and (iii) 1.2 TW/cm$^2$ (for the latter radiance, two runs obtained on different days are plotted together to illustrate the reproducibility). A 1-m long single pass delay line was used to obtain the kinetics of $e_{aq}^-$ on the nanosecond time scale. In (b), the TA signals are normalized at $t=10$ ps. Note the logarithmic scale for the transient absorbance in (a). At high radiance of 400 nm light, the slow decay of the TA kinetics is mainly due to cross recombination of electrons and *OH* radicals and hydronium ions in the water bulk (see Fig. 2(b) for simulations and section 3.4 for more discussion).

**Fig. 8S.**

(a) *Open circles:* Pump-probe TA kinetics of hydrated electron in picosecond (5 ps FWHM) pulse radiolysis of $D_2O$ (1000 nm probe, 1 cm cell). Plotted from the data obtained, with permission, from J. R. Wishart of BNL (ref. 32). The solid line is exponential fit of the fast component.

(b) Temperature dependence of the absorptivity of hydrated electron in $H_2O$ at (i) 820 nm and (ii) 1000 nm. The value of 21000 $M^{-1}$ $cm^{-1}$ is taken for the unit of molar absorptivity. As the temperature of the spur rapidly decreases on the picosecond time scale the absorptivity of hydrated electron in the near IR slowly decreases over several tens of picoseconds. Since the heat diffusivity is more than 10 times greater than the electron diffusivity and the heat deposition occurs only along the thermalization path of the electron, the "cooling" occurs much faster than the geminate recombination.

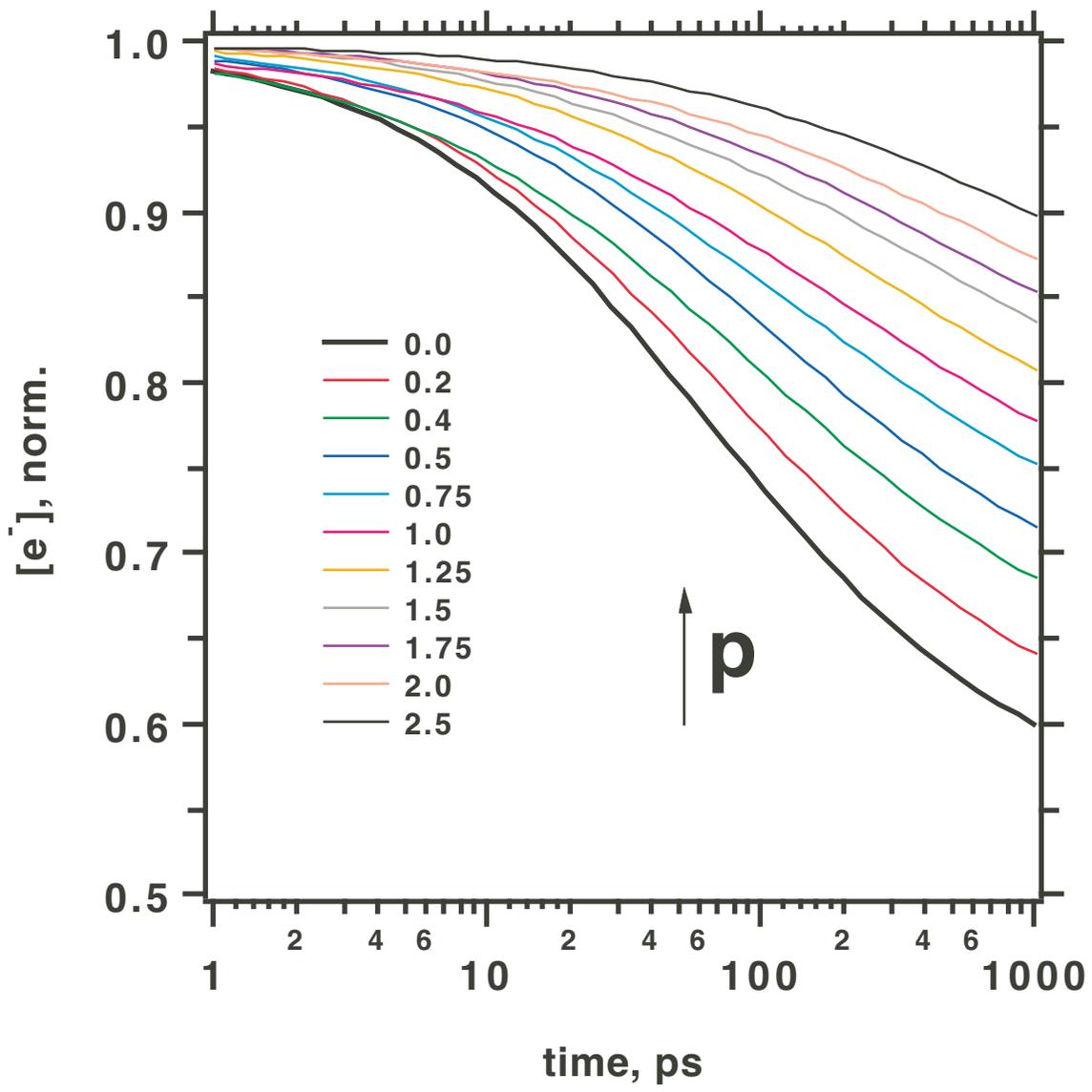

Fig. 1S; Crowell et al.

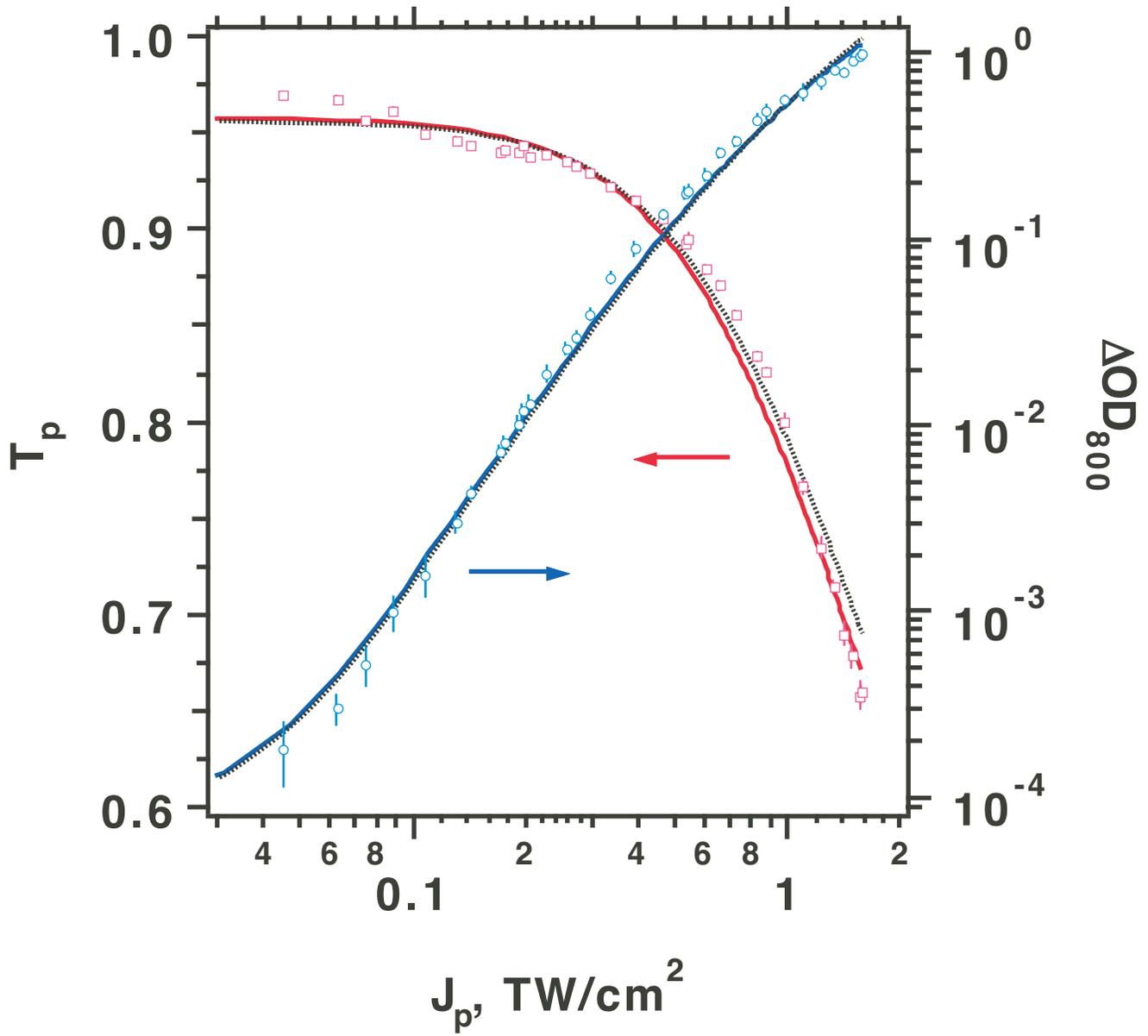

Fig. 2S; Crowell et al.

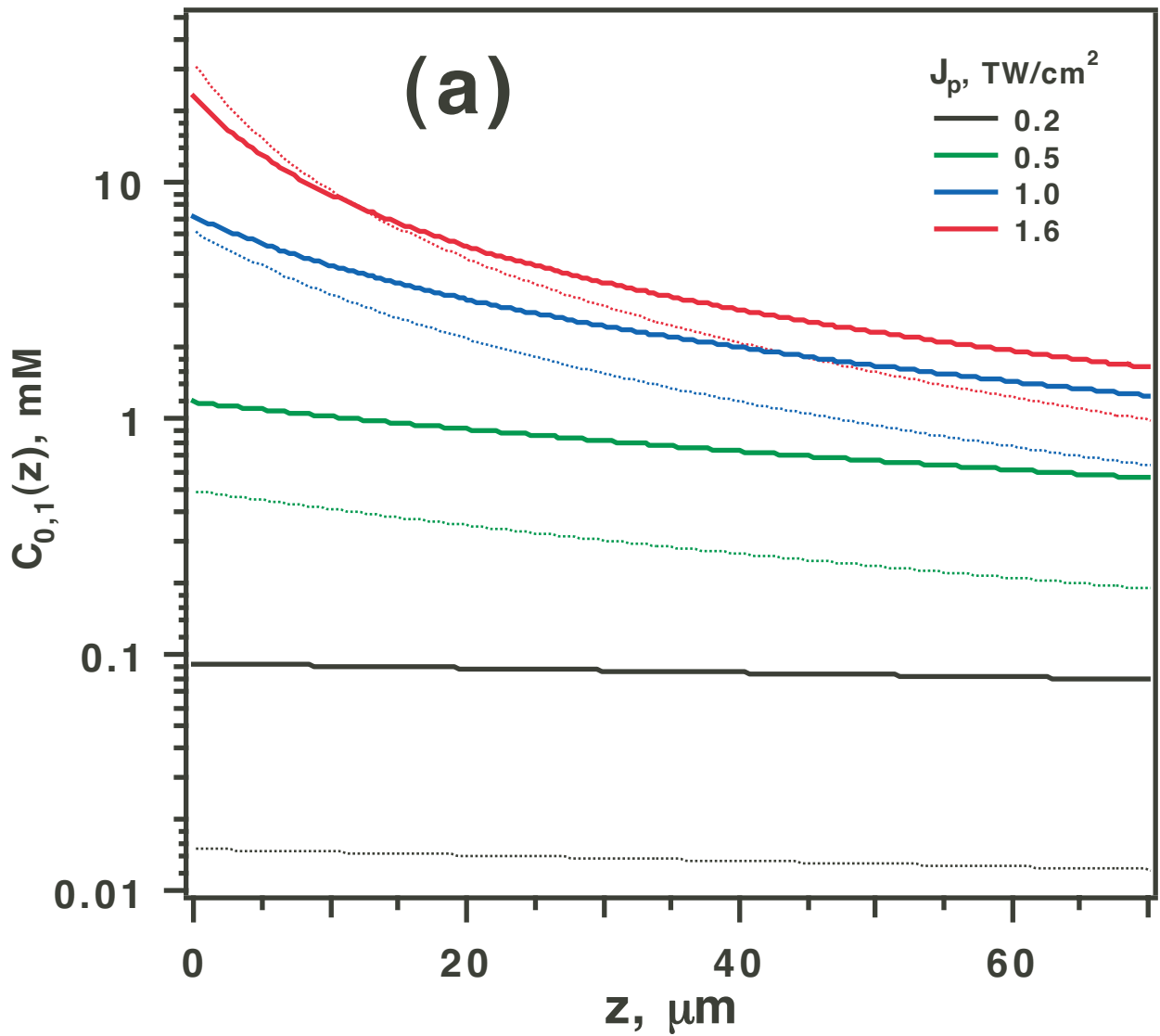

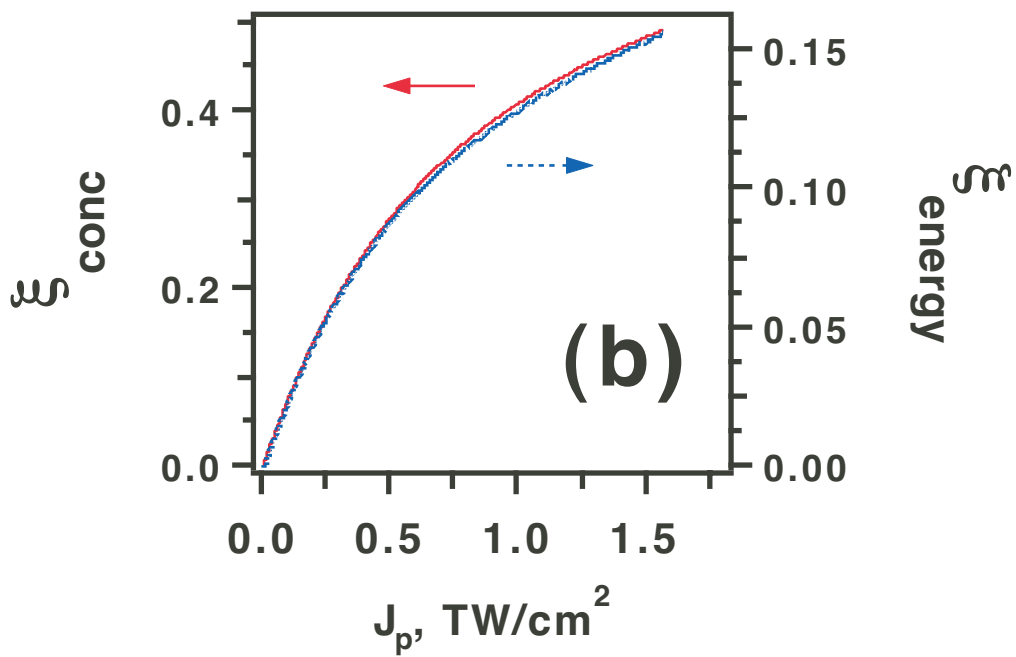

**Fig. 3S; Crowell et al.**

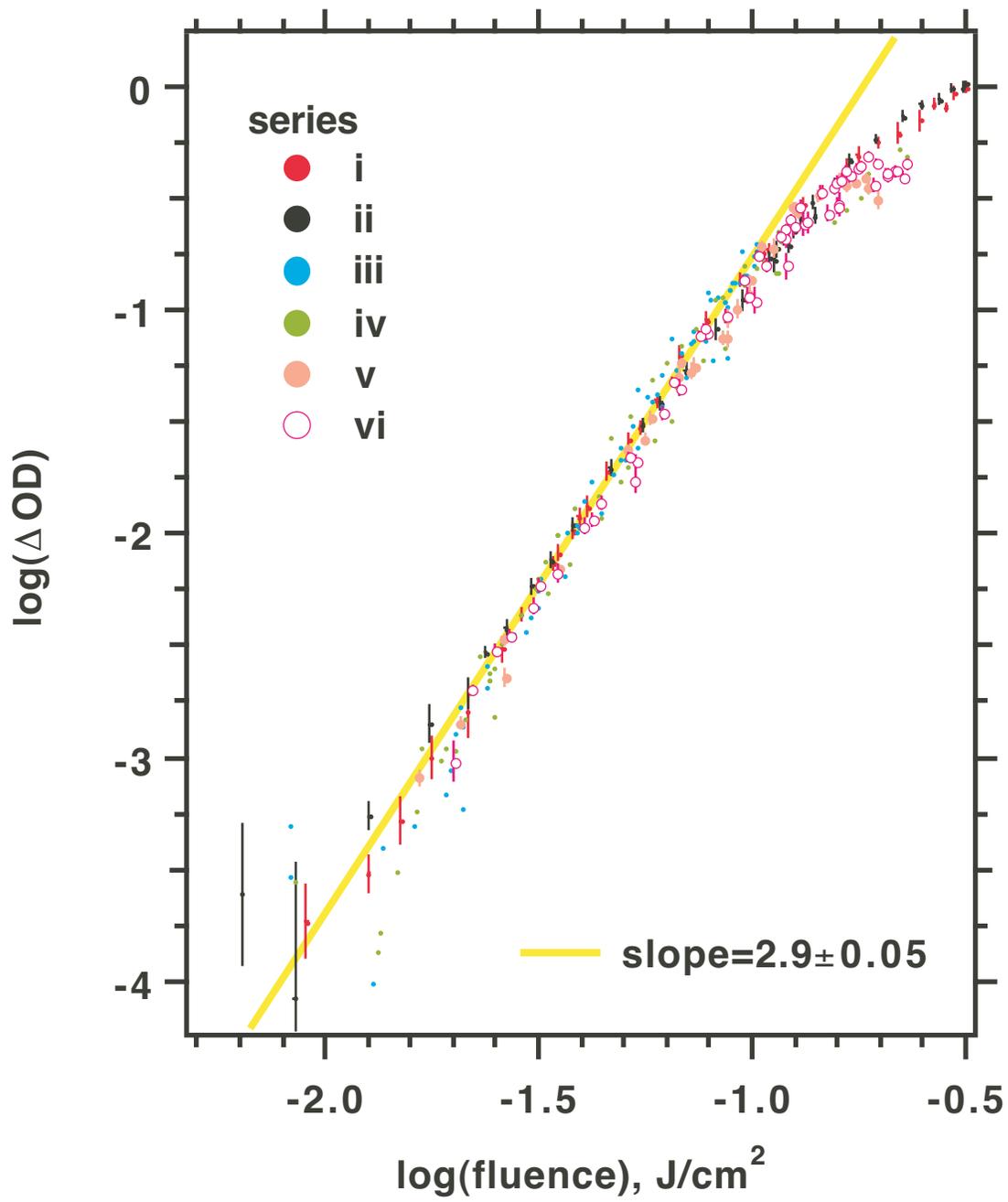

Fig. 4S; Crowell et al.

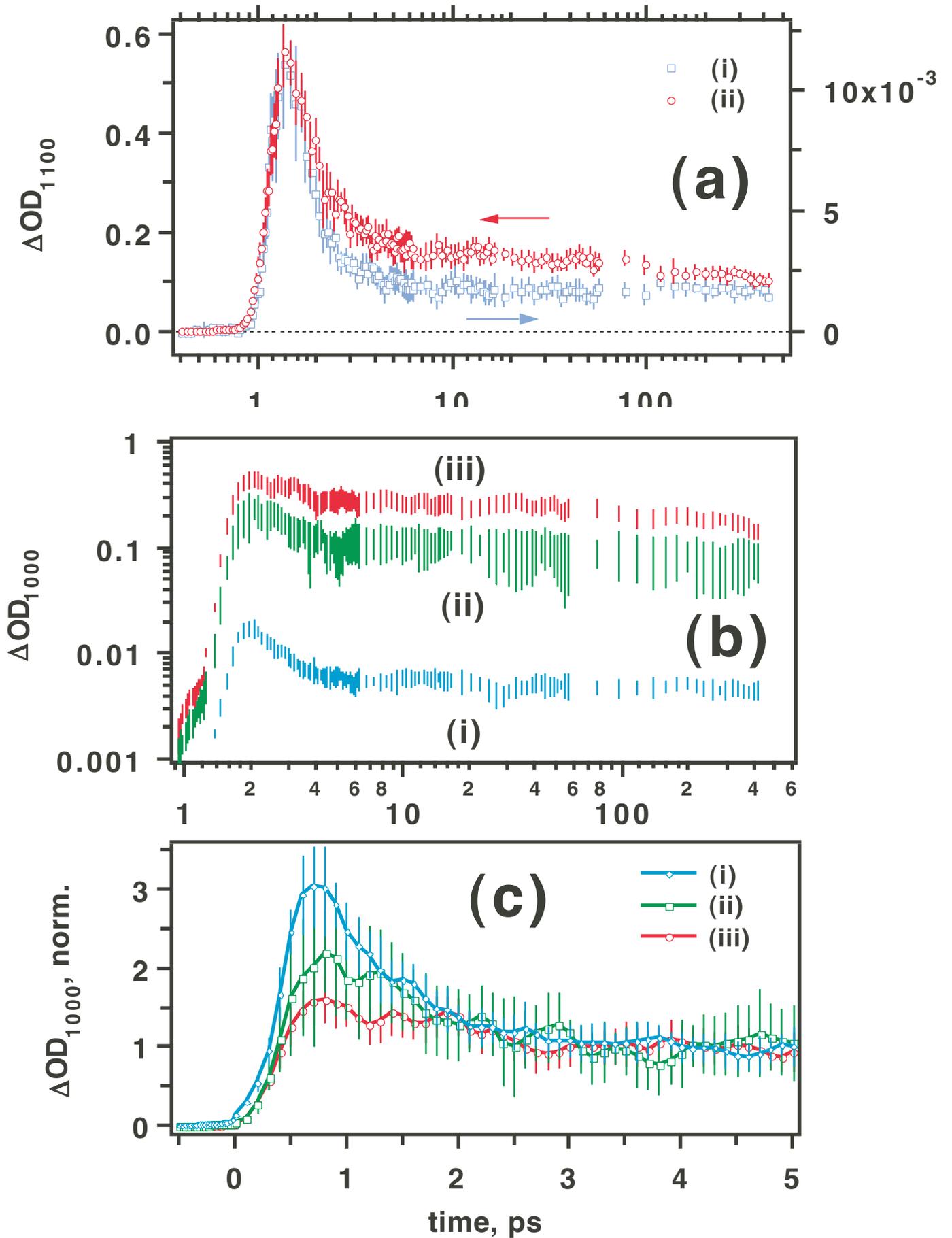

Fig. 5S; Crowell et al.

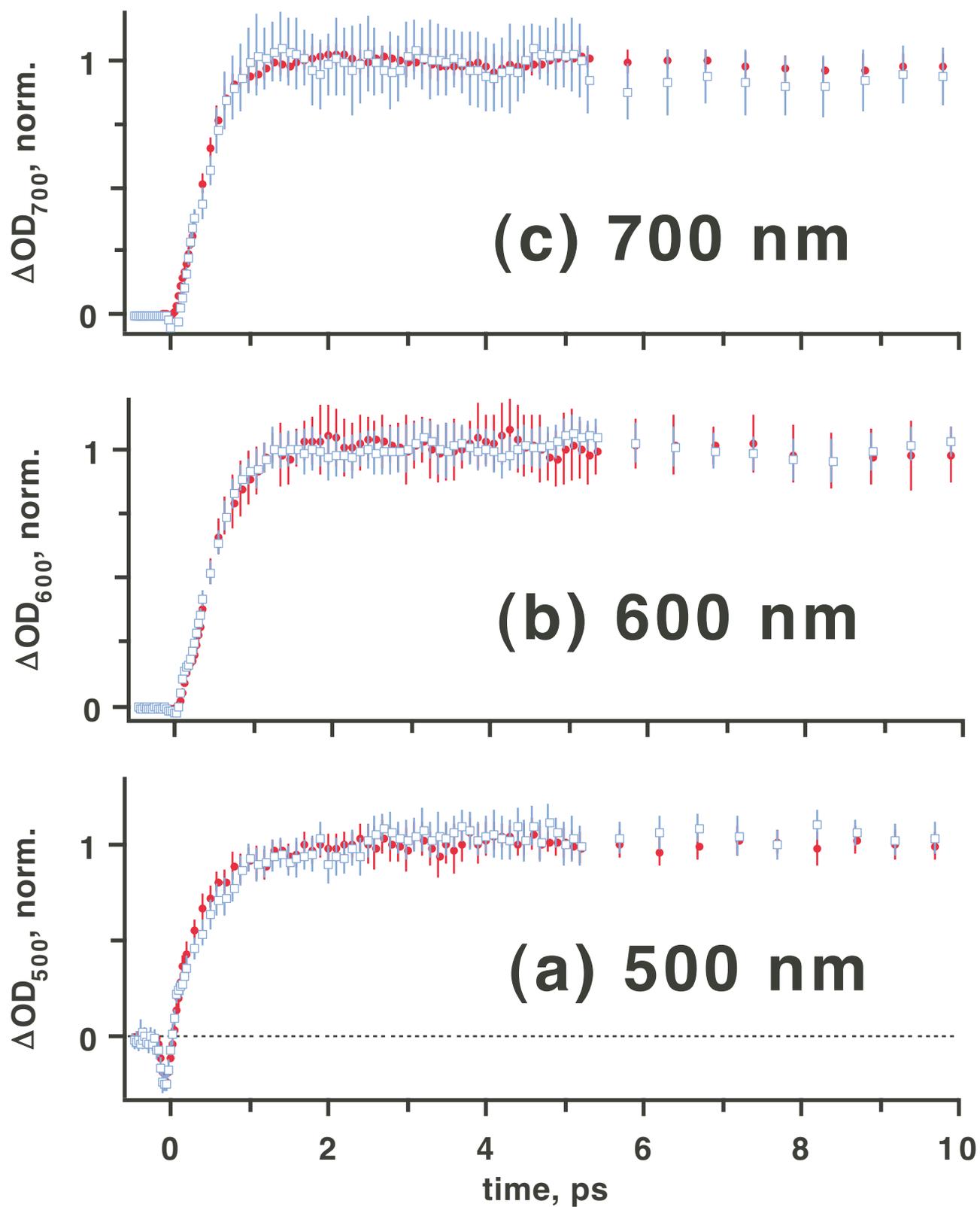

Fig. 6S; Crowell et al.

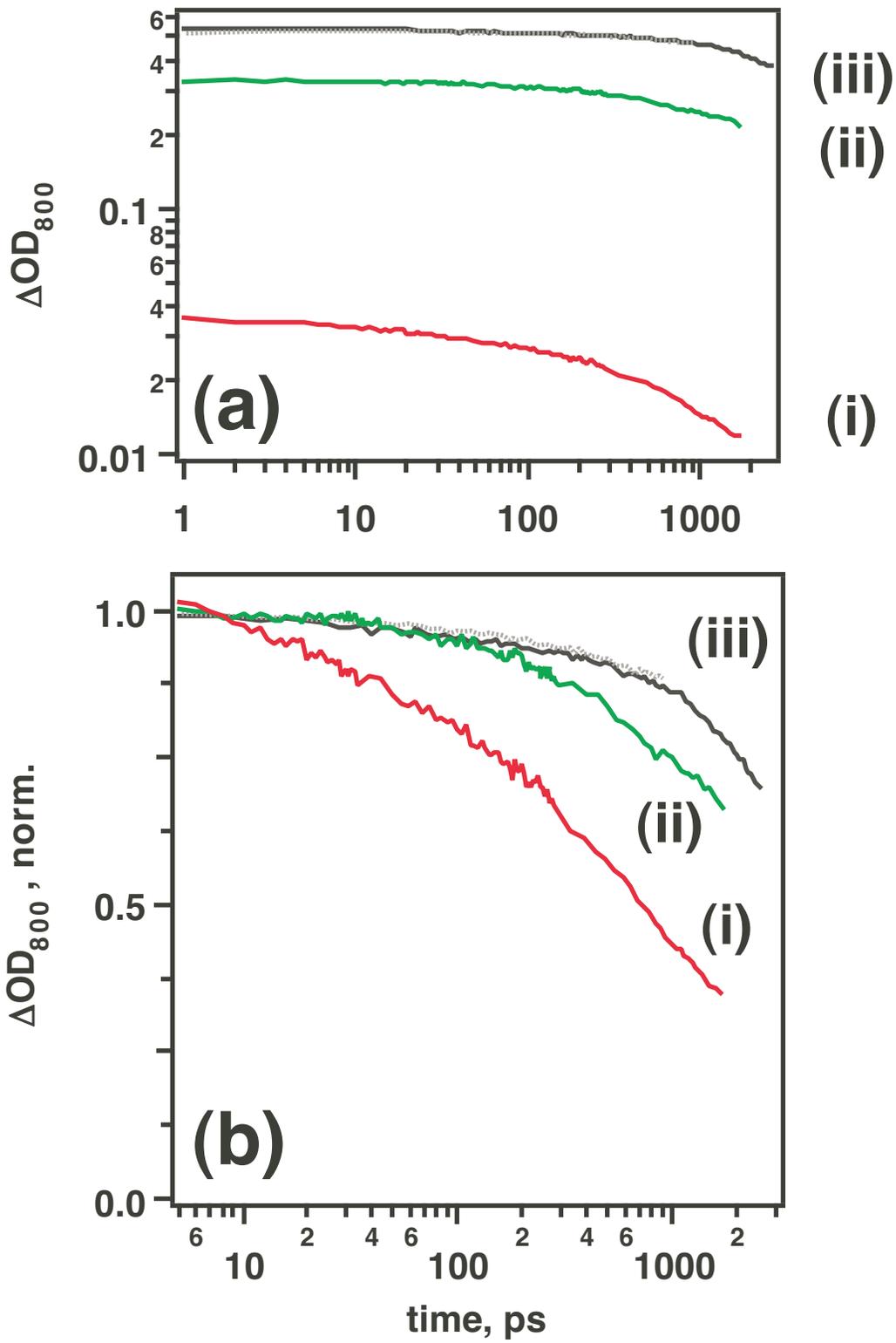

Fig. 7S; Crowell et al.

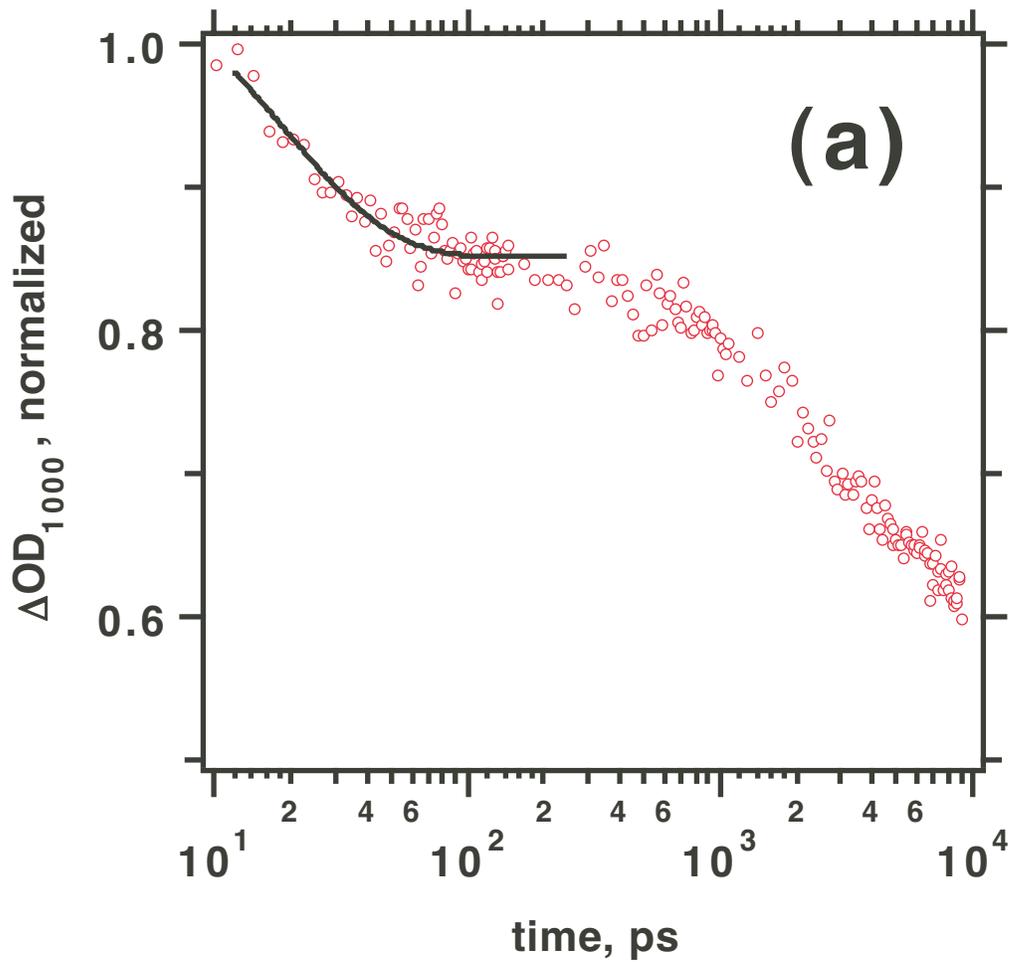

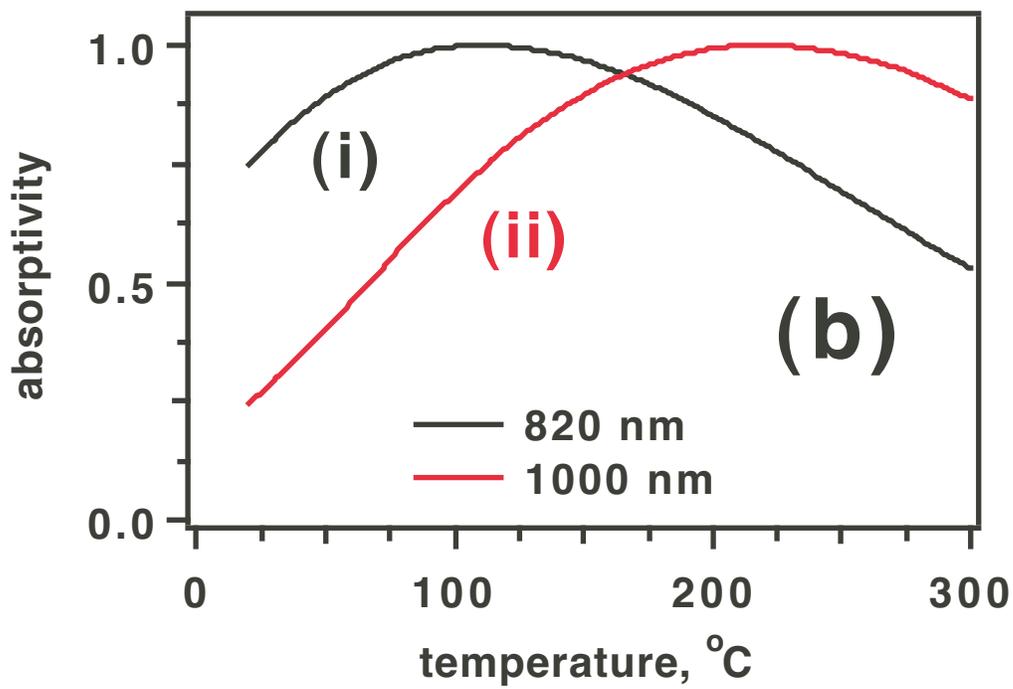

Fig. 8S; Crowell et al.